\documentclass[ aps,twocolumn, prd,showpacs,preprintnumbers,nofootinbib,floatfix]{revtex4-1}
\renewcommand\d{\partial}

\usepackage{amsmath,amssymb,bm,comment,color, graphicx}
\usepackage{hyperref}

\newcommand{\arXiv}[1]{\href{http://www.arXiv.org/abs/#1}{#1}}

\newcommand{\ddd}{\displaystyle}
\newcommand{\lbr}{\left(}
\newcommand{\rbr}{\right)}

\newcommand{\CS}{S}

\def\d{\partial}

\def\half{\frac{1}{2}}

\def\nn{\nonumber}

\advance\voffset by 0.5cm

\makeatletter
\DeclareRobustCommand*{\bfseries}{%
  \not@math@alphabet\bfseries\mathbf
  \fontseries\bfdefault\selectfont
  \boldmath
}
\makeatother

\begin{document}

\preprint{DESY 12-042}
\preprint{ZMP-HH/12-4}

\title{Anisotropic hydrodynamics, holography and the chiral magnetic effect}

\author{Ilmar Gahramanov}
\author{Tigran Kalaydzhyan}
\author{Ingo Kirsch  \medskip}

\affiliation{DESY Hamburg, Theory Group, Notkestrasse 85, D-22607 Hamburg, Germany}
\affiliation{Zentrum f\"{u}r Mathematische Physik, Universit\"{a}t Hamburg, Bundesstrasse 55, D-20146 Hamburg}
\date{\today}
\begin{abstract}
We discuss a possible dependence of the chiral magnetic effect (CME) on the elliptic flow coefficient~$v_2$. We first study this in
a hydrodynamic model  for a static anisotropic plasma with multiple anomalous
$U(1)$ currents. In the case of two charges, one axial and one vector, the CME formally appears as a first-order transport coefficient
in the vector current. We compute this transport coefficient and show its dependence on $v_2$. We also determine
the CME coefficient from first-order corrections to the dual anti-de Sitter background using the fluid-gravity duality. For small anisotropies,
we find numerical agreement with the hydrodynamic result.
\end{abstract}
\maketitle

\section{Introduction}

In the last couple of years the chiral magnetic effect (CME) has attracted much attention
as a candidate for the explanation of an experimentally observed charge
asymmetry in heavy-ion collisions, as seen by the STAR~\cite{STAR}, PHENIX
\cite{PHENIX} and ALICE~\cite{ALICE} collaborations.
The CME is a hypo\-the\-tical phenomenon which states
that, in the presence of a magnetic field $\vec  B$,
an electric current is generated  along $\vec B$  in the background of topologically nontrivial
gluon fields~\cite{Fukushima, Kharzeev}.  Analogous effects were found
earlier in neutrino~\cite{Vilenkin}, electroweak
\cite{Giovannini:1997gp} and condensed matter physics~\cite{condmat}.
Lattice QCD results \cite{lattice1, lattice2, lattice3} suggest the existence of
the effect, although the magnitude of the CME-induced charge asymmetry
may be too small to explain the observed charge asymmetry  \cite{Mueller}.

In a recent experiment, the charge separation is  measured as a function of
the elliptic flow coefficient $v_2$ \cite{Wang}. The data is taken
from (rare) Au+Au collisions with $20-40$\% centrality but different $v_2$. In this way
$v_2$ is varied while at the same time the number of participating nucleons (and
therefore the magnetic field) is kept almost constant. The plots in \cite{Wang} suggest
that the charge separation is proportional to $v_2$.  If this holds true, the charge separation will depend
on the event anisotropy.

In this paper we address the question of whether and how the CME depends
on the elliptic flow $v_2$. We study this both in hydrodynamics and in terms of a holographic
gravity dual.
The hydrodynamical approach to the CME and CME-related phenomena was
proposed in \cite{Son,Zahed,Isachenkov,Kirsch11,Zakharov,Pu,Oz}.
There, the CME appears in form of a nonvanishing transport coefficient in
the electric current, $\vec j = \kappa_B \vec B$, which measures the response of the system to an
external magnetic field \cite{Erdmenger, Son}. In \cite{Kirsch11},
the chiral magnetic conductivity in an isotropic fluid was determined as
\begin{align}
 \kappa_B = C \mu_5 \left( 1 - \displaystyle\frac{\mu \rho}{\epsilon + P} \right) .
\end{align}
The first term is the standard term for the CME and depends only on the axial
anomaly coefficient $C$ and the axial chemical potential $\mu_5$. The second
term proportional to the factor $\frac{\rho}{\epsilon + P}$ depends on the
dynamics of the fluid and has a chance to depend on $v_2$ in the anisotropic case.

\begin{figure}[t]
\centering
\includegraphics[width=8.55cm]{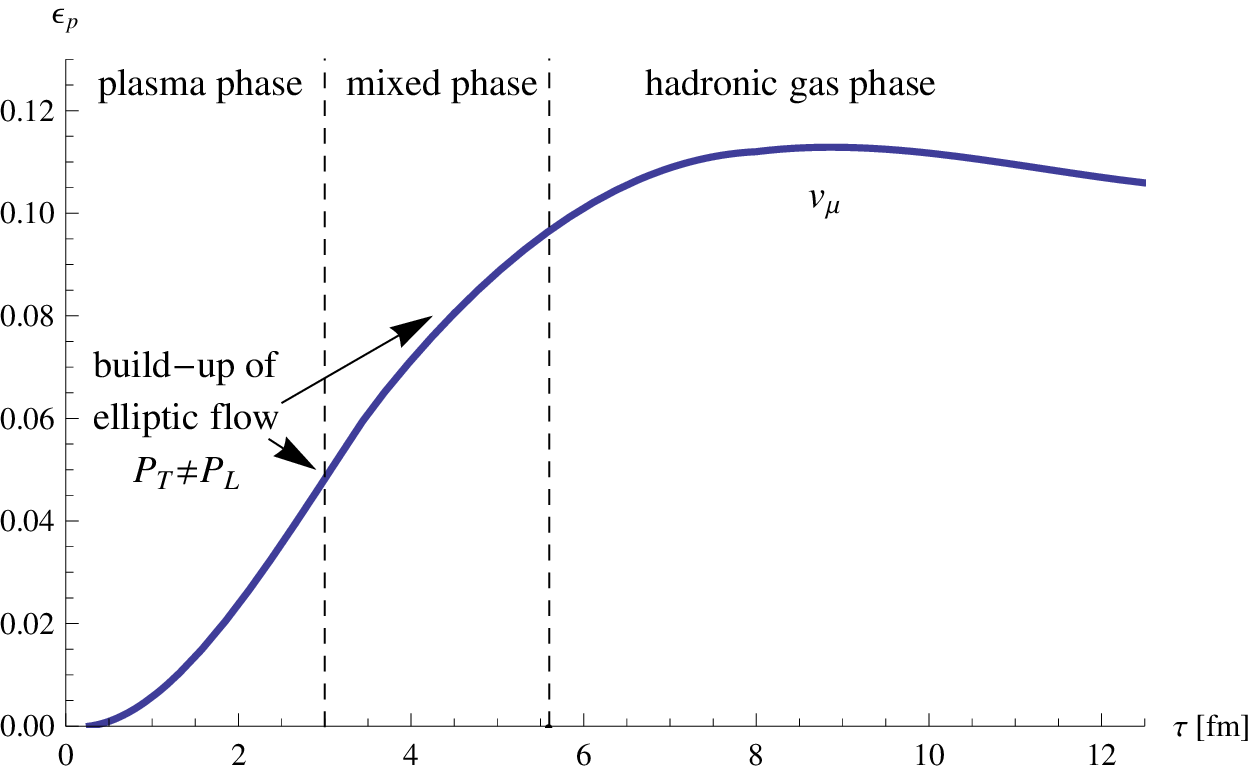}
\vspace{-4.05cm}
\begin{flushright}
\includegraphics[width=4.62cm]{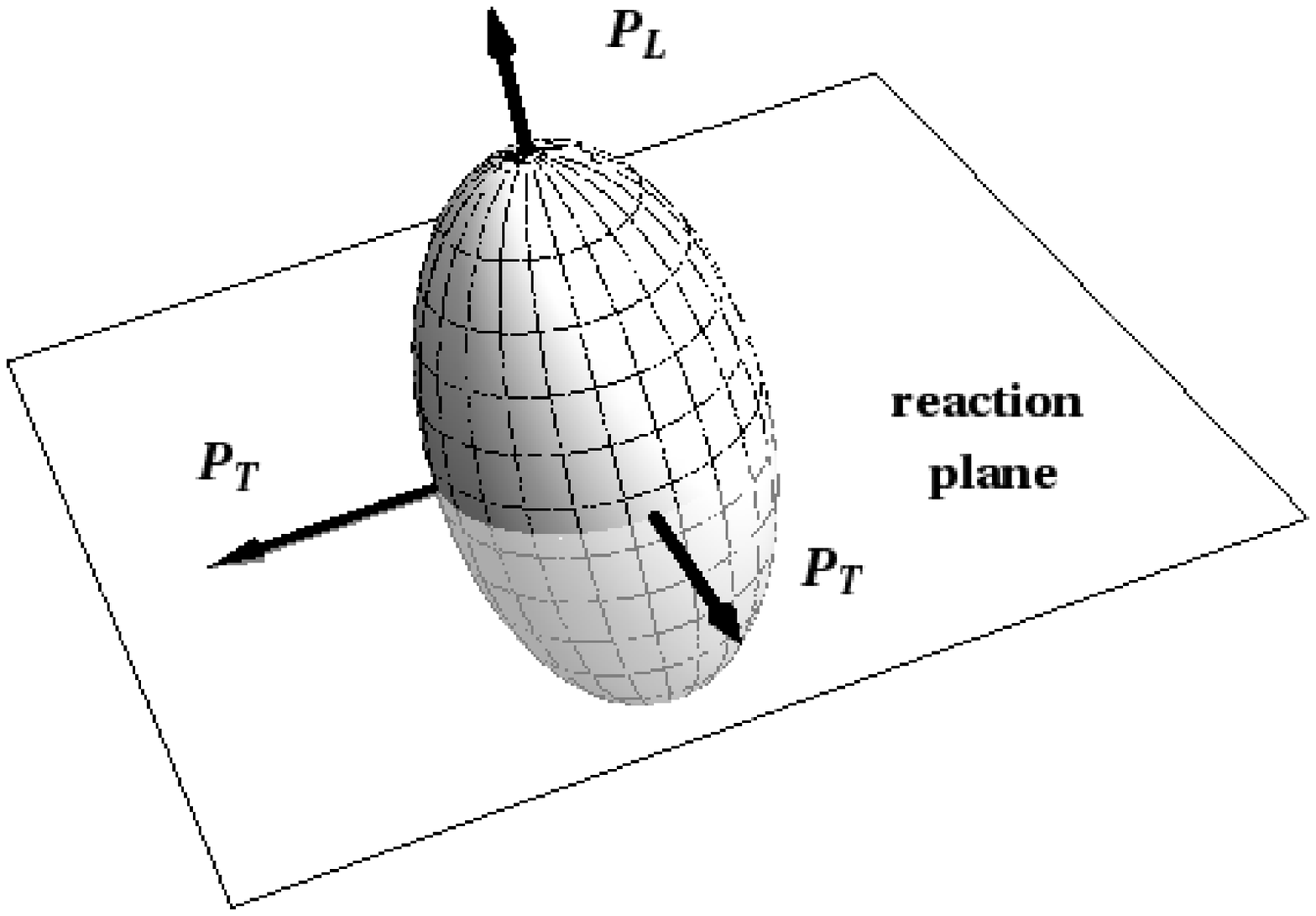}
\hspace{0.1cm}\mbox{}
\end{flushright}
\vspace{0cm}
\caption{\label{epsplot} Sketch of the time evolution of the momentum anisotropy $\varepsilon_p$ (based on~\cite{Huovinen}).
The small figure shows the orientation of $P_L$ and $P_T$ with respect to the reaction plane.}
\end{figure}

In the first part of the paper we study this
in a hydro\-dynamic  model for an {\em anisotropic} fluid with multiple anomalous $U(1)$ charges
(This model extends those in \cite{Ryblewski, Florkowski,
  Ryblewski:2011aq}). We compute the CME coefficient $\kappa_B$ and express the result
in terms of the momentum anisotropy~$\varepsilon_p$  \cite{Kolb} defined as
\begin{align}\label{epsP}
\varepsilon_p = \frac{\langle P_T - P_L \rangle}{\langle P_T + P_L \rangle}\,,
\end{align}
where $P_T$ and $P_L$ are the pressures in the plane transverse to the beam line
(In our conventions the indices $L$ and $T$ refer to the longitudinal and transverse direction
with respect to an anisotropy vector $v_\mu$ normal to the reaction plane,
see Fig.~\ref{epsplot}). A sketch of $\varepsilon_p$  as
a function of the proper time $\tau$ is  shown in Fig.~\ref{epsplot}.
 $\varepsilon_p$ describes the build-up of the elliptic flow in off-central collisions.
Our model describes a state after
thermalization with unequal pressures $P_T \neq P_L$. At freeze-out
$\varepsilon_p$ roughly equals $v_2$, and we find that for small anisotropies
the CME-coefficient $\kappa_B$ increases linearly with $v_2$.

In the second part of the paper we perform a holographic computation of
$\kappa_B$ in the dual gravity model. A  similar computation
was previously done  in \cite{Kirsch11} for the
STU model \cite{Cvetic}, a string-theory-inspired
prototype of an (isotropic) anti-de Sitter (AdS) black hole solution with three $U(1)$ charges.
Other holographic approaches to the CME can be found in
\cite{Lifschytz,Yee_CME,Zayakin,Rubakov,Landsteiner,
  Rebhan,Brits,Landsteiner2,Hoyos:2011us,Bhattacharya:2011tra,Hu:2011ze}.

In the anisotropic case, we first need to construct an appropriate gravity
background. As an ansatz, we choose a multiply charged
AdS black hole solution with some additional functions $w_L$ and $w_T$ inserted
which will make the background anisotropic and $\varepsilon_p$-dependent.
Since analytical solutions for charged anisotropic backgrounds are notoriously
difficult to find, we will use shooting techniques to find a numerical solution.
Other AdS backgrounds dual to anisotropic fluids
are constructed in \cite{Mateos, Erdmenger2011, Witaszczyk}.

As the AdS solution in \cite{Witaszczyk}, the background is
static and does not describe the  process of
isotropization. Even though such models have some limitations~\cite{Witaszczyk},
they are nevertheless useful for the computation of transport coefficients. We
show this, following \cite{Kirsch11}, by determining $\kappa_B$  from the
first-order corrections to this background using the fluid-gravity duality \cite{Hubeny}.
For small anisotropies, we find numerical agreement with the hydrodynamic
result for $\kappa_B$.  Other (dissipative) transport coefficients in strongly-coupled
anisotropic plasmas are discussed in
\cite{Rebhan:2011ke, Rebhan:2011vd, Giataganas:2012zy}.

The paper is organized as follows. In Sec.~\ref{sec2} we review the
hydrodynamics of an anisotropic relativistic fluid with several $U(1)$
charges and triangle anomalies. We then compute the vortical and
magnetic conductivities of such a fluid by extending the method of Son
and Surowka \cite{Son} to the anisotropic case. In Sec.~\ref{sec3}
we construct the dual gravity background and present a numerical
solution for its gauge field and metric functions. In
Sec.~\ref{sec4} we use this background to perform a holographic
computation of the vortical and magnetic conductivities.

\section{Hydrodynamics of anisotropic fluids
with triangle anomalies}\label{sec2}

The hydrodynamic regime of isotropic relativistic fluids with triangle
anomalies has been studied in \cite{Son, Zakharov, Isachenkov, Pu, Oz,
  Zahed}, and much can be taken over to the anisotropic case.  Such
fluids typically contain $n$ anomalous $U(1)$ charges which commute
with each other.  The anomaly coefficients are given by a totally
symmetric \mbox{rank-$3$} tensor~$C^{abc}$. The hydrodynamic equations
are
\begin{align}\label{hydroeqn}
  \partial_\mu T^{\mu\nu} = F^{a\nu\lambda} j^{a}_\lambda
  \,,\qquad
  \d_\mu j^{a\mu} = C^{abc} E^b\cdot B^c \,,
\end{align}
where $E^{a\mu}=F^{a \mu\nu} u_\nu$, $B^{a\mu} =
\frac{1}{2}\epsilon^{\mu\nu\alpha \beta}u_\nu F^a_{\alpha\beta}$
($a=1,...,n$) are electric and magnetic fields, and
$F^a_{\mu\nu}=\partial_\mu A^a_\nu-\partial_\nu A^a_\mu$ denotes the
gauge field strengths. As in \cite{Son}, we expand the constitutive equations
for $T^{\mu\nu}$ and $j^\mu$ up to first order, taking  $A^a_\mu \sim O(p^0)$
and $F^a_{\mu\nu} \sim O(p)$. The gauge fields $A^a_\mu$ are nondynamical.

In {\em anisotropic} relativistic fluids, the hydrodynamic equations
are again given by (\ref{hydroeqn}) but the stress-energy tensor $T^{\mu\nu}$ and $U(1)$
currents $j^{a\mu}$ now have the more general form\footnote{The
  symmetries allow in principle for more general currents $j^{a\mu} =
  \rho^a u^\mu + c^a v^\mu + \nu^{a \mu}$ with some coefficients
  $c^a$. Here we switch off all the `electric' background currents,
  $c^a =0$.}
\begin{align}
  T^{\mu\nu}&=(\epsilon+P_T)u^\mu u^\nu+P_T g^{\mu \nu}-\Delta v^\mu
  v^\nu+\tau^{\mu \nu}
  \,,\label{AT}\\
  j^{a\mu} &= \rho^a u^\mu + \nu^{a \mu}\,, \label{currents}
\end{align}
where $\epsilon$ is the energy density, $\rho^a$ are the $U(1)$ charge
densities, $\Delta=P_T-P_L$, and $P_T$ and $P_L$ denote the transverse
and longitudinal pressures, respectively \cite{Ryblewski, Florkowski,
  Ryblewski:2011aq}. $g_{\mu\nu}$ is the metric with signature
$(-,+,+,+)$.  $\tau^{\mu\nu}$ and $\nu^{a \mu}$ denote higher-gradient
corrections, for which we require $u_\mu \tau^{\mu\nu}=0$ and $u_\mu
\nu^{a \mu}=0$.

The four-vectors $u^\mu$ and $v^\mu$ describe the flow of the fluid
and the direction of the longitudinal axis, respectively.  The vector
$v^\mu$ is spacelike and orthogonal to $u^\mu$,
\begin{eqnarray}
  u_\mu u^\mu=-1 \,,\qquad
  v_\mu v^\mu=1 \,,\qquad
  u_\mu v^\mu=0\,.
\end{eqnarray}
It is convenient to define the proper time $\tau$ by $\partial^\nu \ln
\tau \equiv v^\mu\partial_\mu v^\nu$ \cite{Florkowski}. In the rest
frame of the fluid, $u^\mu=(1,0,0,0)$ and $v^\mu=(0,0,0,1)$, the
stress-energy tensor becomes diagonal,
\begin{equation}\label{diagTmn}
  T^{\mu \nu} =  \left(
\begin{array}{cccc}
  \epsilon & 0 & 0 & 0 \\
  0 & P_T & 0 & 0 \\
  0 & 0 & P_T & 0 \\
  0 & 0 & 0 & P_L
\end{array} \right).
\end{equation}
In conformal fluids, the stress-energy tensor is traceless,
$T^{\mu}{}_\mu=0$, and $\epsilon = 2P_T+P_L$. Clearly, the
isotropic case corresponds to equal pressures $P_T$ and $P_L$,
$P=P_T=P_L$.

\medskip

For simplicity, we restrict to the case of a single charge in
Secs.~\ref{sec2A} and \ref{sec2B}, $n=1$.  In Secs.~\ref{sec2C}
and \ref{sec2D} we generalize our findings to arbitrary $n$ and
discuss the case $n=2$, which is relevant for the CME.

\subsection{Thermodynamics of an anisotropic fluid with chemical
  potential \texorpdfstring{$(n=1)$}{}}\label{sec2A}

Hydrodynamic models for an anisotropic fluid (without chemical
potential) have been studied in
 \cite{Ryblewski, Florkowski,
  Ryblewski:2011aq}. Following these works, we
derive some thermodynamic identities, now for the case of a fluid with
a chemical potential~$\mu$.

These identities can be found by computing the quantity $I_0=u_\nu \partial_\mu
T^{\mu\nu}+\mu \partial_\mu j^\mu$ at zeroth order. Since the right-hand side
of (\ref{hydroeqn}) can be dropped at order $O(p^0)$, we have $I_0=0$.
Using $\partial_\mu (s u^\mu)=0$, we get
\begin{align}
  u_\nu \partial_\mu T^{\mu\nu}&= - u^\mu \partial_\mu \epsilon -
  (\epsilon+ P_T) \partial_\mu u^\mu - \Delta u_\nu \partial^\nu
  \ln \tau
  \nonumber\\
  &=- u^\mu \partial_\mu \epsilon + \frac{\epsilon+P_T}{s}
  u^\mu \partial_\mu s - \frac{\Delta}{\tau} u^\mu \partial_\mu \tau \,,
  \label{dTmunu}\\
  \mu \partial_\mu j^\mu & = \mu
  (\partial_\mu \rho) u^\mu - \frac{\mu \rho}{s} u^\mu \partial_\mu
  s\,.
\end{align}
As in \cite{Florkowski}, we consider a generalized energy density
$\epsilon = \epsilon (s, \rho, \tau)$, which depends not only on
the entropy density~$s$ and particle density~$\rho$ but also on the
new variable $\tau$. Its differential is
\begin{align} \label{deps}
  d \epsilon = \left(\frac{\partial \epsilon}{\partial
      s}\right)_{\rho,\tau} ds + \left( \frac{\partial
      \epsilon}{\partial \rho } \right)_{s,\tau} d\rho + \left(
    \frac{\partial \epsilon}{\partial \tau}\right)_{s,\rho} d\tau
  \,,
\end{align}
with
\begin{align}
  \left(\frac{\partial \epsilon}{\partial s}\right)_{\rho,\tau}
  \!\!= T \,,\quad \left( \frac{\partial \epsilon}{\partial \rho }
  \right)_{s,\tau} \!\!=\mu \,,\quad \left( \frac{\partial
      \epsilon}{\partial \tau}\right)_{s,\rho} \!\! =-
  \frac{\Delta}{\tau}\,.
\end{align}
The temperature and the chemical potential are defined in the usual
way. If we also impose $ \left( {\partial \epsilon}/{\partial
    \tau}\right)_{s,\rho} =- {\Delta}/{\tau}$ and substitute
(\ref{deps}) into (\ref{dTmunu}), then $I_0=0$ implies the following
thermodynamical identities for an anisotropic fluid:
\begin{align}
  \epsilon + P_T &= Ts + \mu \rho \,,\label{thermodyn1} \\
  d P_T &= \frac{\Delta}{\tau} d\tau + s dT + \rho d\mu \,, \label{thermodyn2}\\
  d\epsilon &= Tds + \mu d\rho - \frac{\Delta}{\tau}d\tau
  \,,\label{thermodyn3}
\end{align}
in agreement with \cite{Florkowski} for $\mu=0$.

\subsection{Vortical and magnetic coefficients  \texorpdfstring{$(n=1)$}{}}\label{sec2B}

We now discuss corrections to the $U(1)$ current \mbox{$j^\mu\equiv
  j^{1\mu}$} $(n=1)$. In anisotropic fluids the transport coefficients
are usually promoted to tensors such that one should consider
first-derivative corrections of the type
\begin{align}
  \nu^\mu &= (\xi_\omega)^\mu{}_\nu  \omega^\nu + (\xi_B)^\mu{}_\nu B^\nu \,,
\end{align}
where $\omega^\mu=\textstyle
  \half\epsilon^{\nu\rho\sigma\mu}u_\nu\partial_\rho u_\sigma$ is the vorticity, and $B^\mu$ is an external
magnetic field. In Landau frame $u_\mu \nu^\mu = 0$ and therefore
$u_\mu (\xi_{\omega})^\mu{}_\nu \omega^\nu=0$ (and similar for
$(\xi_{B})^\mu{}_\nu$).  This is satisfied {\em e.g.}\ for
$(\xi_\omega)^\mu{}_\nu = \xi_\omega \delta^\mu{}_\nu$, since $ u_\mu
\omega^\mu = 0$ (We do not consider other components of $\xi_\omega$
here). We therefore restrict to consider corrections of the type
\begin{align}
  \nu^\mu &= \xi_\omega  \omega^\mu + \xi_B B^\mu \,,
\end{align}
as in the isotropic case \cite{Son}.  Our goal is to compute the
vortical and magnetic conductivities $\xi_\omega$ and $\xi_B$.  These
transport coefficients can be found by assuming the existence of an
entropy current $s^\mu$ with a non-negative derivative, $\partial_\mu
s^\mu \geq 0$. The computation closely follows that of~\cite{Son}.

The hydrodynamic Eqs.~(\ref{hydroeqn}) imply that the quantity
\begin{equation}
  I_1 = u_\nu \partial_\mu T^{\mu\nu}+\mu\partial_\mu j^\mu
  + E^\mu \nu_\mu-\mu C E^\mu B_\mu
\end{equation}
vanishes at first order, $I_1=0$. Substituting the explicit
expressions for the stress-energy tensor and $U(1)$ currents into
$I_1$ and using the thermodynamical identities (\ref{thermodyn1}) and
(\ref{thermodyn3}), we find
\begin{align}
  \partial_\mu \left(su^\mu-\frac\mu T\nu^\mu\right) & = -\frac{1}{T}
  \partial_\mu u_\nu\tau^{\mu\nu} - \nu^\mu\left(
  \partial_\mu\frac\mu T - \frac{E_\mu}{T} \right) \nonumber\\
  &~~~- C \frac\mu T E\cdot B \,,  \label{entprod}
\end{align}
which is exactly the same equation for the entropy production as in
the isotropic case \cite{Son}.

In the following, we will need  the identities
\begin{align}
  \label{Paromega} \partial_{\mu}\omega^{\mu} & =
  -\frac{2}{\epsilon+P_T}\omega^\mu(\partial_{\mu}P_T
  -{\Delta \partial_\mu \ln \tau} -\rho E_\mu)\,, \\
  \label{ParB} \partial_{\mu}B^{\mu} & =
  -2\omega^{\mu}E_{\mu}-\frac{B^\mu}{\epsilon+P_T} (\partial_{\mu}P_T-
  \Delta \partial_\mu \ln \tau-\rho E_\mu) \,, \nonumber
\end{align}
which we derived from ideal hydrodynamics in Appendix~\ref{appA}.  In deriving
these identities we assumed that the fluid satisfies
\begin{align}
  \partial_\mu v^\mu=0 \,,\qquad v^\mu \partial_\mu \Delta=0 \,.
\end{align}
The first equation is basically a ``continuity equation'' for the vector
$v^\mu$. There are no sources for the generation of anisotropy. The
second equation imposes an orthogonality relation between the gradient
of the pressure difference $\Delta=P_T-P_L$ and $v^\mu$.

As in \cite{Son}, we assume a generalized entropy current of the form
\begin{align}
  s^\mu &= su^\mu - \frac\mu T\nu^\mu + D\omega^\mu + D_B B^\mu,
\end{align}
where $\xi_\omega$, $\xi_B$, $D$, and $D_B$ are functions of $T$,
$\mu$ and $\tau$.  We now compute $\partial_\mu s^\mu$, using
(\ref{entprod}) and (\ref{Paromega}) and impose $\partial_\mu s^\mu
\geq 0$. Since the coefficients in front of $\omega^\mu$, $B^\mu$,
$\omega_\mu E^\mu$ and $ E_\mu B^\mu$ inside $\partial_\mu s^\mu$ can
have either sign, we require them to vanish and obtain the following
four differential equations:
\begin{align}
  &\partial_\mu D - \frac{2D}{\epsilon+P_T}(\partial_\mu P_T -\Delta
  \partial_\mu \ln \tau) - \xi_\omega\partial_\mu \frac \mu T = 0\,,
    \label{eq1}\\
  &\partial_\mu D_B - \frac{D_B}{\epsilon+P_T}(\partial_\mu P_T -\Delta \partial_\mu \ln \tau )- \xi_B\partial_\mu
    \frac \mu T = 0\,, \label{eq2} \\
  &\frac{2\rho D}{\epsilon+P_T}-2D_B + \frac{\xi_\omega}{ T} = 0\,, \label{eq3} \\
  & \frac{\rho D_B}{\epsilon+P_T} + \frac{\xi_B}T - C \frac\mu T =0\,. \label{eq4}
\end{align}
For $\Delta=0$, these equations reduce to those in the isotropic case
\cite{Son}.

In Appendix~\ref{appB} we solve (\ref{eq1})--(\ref{eq4}) for $D$, $D_B$,
$\xi_\omega$ and~$\xi_B$.  As a result, we find the vortical and
magnetic conductivities
\begin{align}
  \nonumber \xi_\omega &= C \left(\mu^2 - \frac{2}{3}
    \frac{\rho \mu^3}{\epsilon+P_T}\right)
  + {\cal O}(T^2)\,,\\
  \xi_B &= C \left(\mu-\frac{1}{2}\frac{\rho
      \mu^2}{\epsilon+P_T}\right) + {\cal O}(T^2) \,,
\end{align}
where ${\cal O}(T^2)$ denotes terms proportional to $T^2$, see
(\ref{resxi}) in Appendix~\ref{appB}. These terms are related to
gravi\-ta\-tional triangle anomalies \cite{Oz, Landsteiner:2011cp} and may, in
the anisotropic case, depend on the proper time $\tau$. In the absence
of gravitational anomalies, which we do not discuss in this paper, the
conductivities do not depend on $\tau$.  Apart from these changes in
${\cal O}(T^2)$, the relations have the same form as in the isotropic
case but with $P$ replaced by the transverse pressure $P_T$.

\subsection{Multiple charge case (\texorpdfstring{$n$}{} arbitrary)}\label{sec2C}

The generalization of the previous computation to a fluid with
multiple anomalous $U(1)$ charges is straightforward, and we only
state the result here. The corrections $\nu^{a\mu}$ of the currents
$j^{a\mu}$ in (\ref{currents}) are
\begin{align}
  \nu^{a\mu} &= \xi^a_\omega  \omega^\mu + \xi^{ab}_B  B^{b\mu} \,,
\end{align}
with [terms of order ${\cal O}(T^2)$ ignored]
\begin{align}
\xi^a_\omega &= C^{abc} \mu^b\mu^c
    - \frac{2}{3} \rho^a C^{bcd} \frac{\mu^b\mu^c\mu^d}{\epsilon+P_T}\label{xi} \,,\\
 \xi_B^{ab} &=  C^{abc} \mu^c
    - \frac{1}{2}\rho^a C^{bcd} \frac{\mu^c\mu^d}{\epsilon+P_T} \label{xiB} \,.
\end{align}
These are simple generalizations of the corresponding conductivities
in the isotropic case \cite{Son, Oz}.

\subsection{Chiral magnetic and vortical effect \texorpdfstring{$(n=2)$}{}}\label{sec2D}

Physically, the most interesting case is that involving two charges
($n=2$) \cite{Isachenkov,Pu, Kirsch11}. The chiral magnetic effect \cite{Kharzeev} can be described
by one axial and one vector $U(1)$, denoted by $U(1)_A \times U(1)_V$.
A convenient notation for the gauge fields and currents is
($a,b,...=1,2$)
\begin{align}
{A}_\mu^A &= {A}^1_\mu \,,\qquad {A}_\mu^V = {A}^2_\mu \,,\nonumber\\
j^{\mu}_5&=j^{1\mu}  \,, \qquad j^{\mu}= j^{2\mu} \,.
\end{align}

\begin{figure}[t]
\centering
\includegraphics[angle=0, width=8cm]{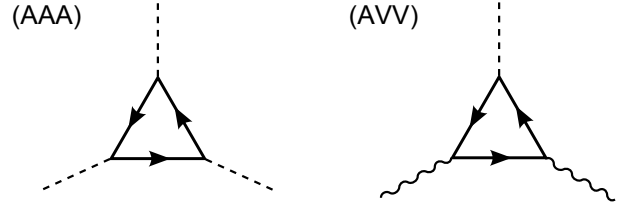}
\caption{Anomalous diagrams corresponding to $C^{111}$(left) and to
  $C^{122}=C^{221}=C^{212}$ (right).  Dashed (wavy) lines denote the axial (vector)
  currents/fields.}\label{fig2}
\end{figure}

Let us now derive the chiral magnetic and vortical effects from
(\ref{xi}) and (\ref{xiB}).  $C-$parity allows for two anomalous
triangle diagrams, (AAA) and (AVV), shown in Fig.~\ref{fig2}, while
diagrams of the type (VVV) and (VAA) vanish.  Accordingly, the anomaly
coefficients are
\begin{align}
C^{121}&=C^{211}=C^{112}=0\,, \quad &(VAA)\nonumber\\
C^{222}&=0 \,, \nonumber \quad &(VVV) \nonumber\\
C^{111}&\neq 0 \,, \quad &(AAA) \nonumber\\
C^{122}&=C^{221}=C^{212}\neq 0 \,. \quad &(AVV)
\end{align}
The hydrodynamic Eqs.~(\ref{hydroeqn}) then imply nonconserved
vector and axial currents
\begin{align}
 \partial_\mu j^\mu &= \textstyle -\frac{1}{4} (C^{212} F^A_{\mu\nu} \tilde F^{V\mu\nu}
 + C^{221} F^V_{\mu\nu} \tilde F^{A\mu\nu}) \,, \nonumber\\
 \partial_\mu j^\mu_5 &=  \textstyle -\frac{1}{4}(C^{111} F^A_{\mu\nu} \tilde F^{A\mu\nu}
+ C^{122} F^V_{\mu\nu}\tilde F^{V\mu\nu}) \,,
\end{align}
where we rewrote $E^b \cdot B^c = -\frac{1}{4}F^b_{\mu\nu}\tilde
F^{c\,\mu\nu}$ (with $\tilde F^{a \mu\nu}=\frac{1}{2}\varepsilon^{\mu\nu\rho\sigma}F^a_{\rho\sigma}$).

To restore conservation of the vector current, we add the
(topological) Bardeen term to the boundary theory,
\begin{align}
  S_B = c_B \int d^4 x\, \epsilon^{\mu\nu\lambda\rho}A_\mu^A A_\nu^V
  F^V_{\lambda\rho} \,.
\end{align}
Combining the corresponding Bardeen currents
\begin{align}
  j^\mu_B &= c_B \varepsilon^{\mu\nu\lambda\rho} (A^V_\nu
  F_{\lambda\rho}^A
  - 2 A_\nu^A F^V_{\lambda\rho} )\,, \nonumber \\
  j^\mu_{5,B} &= c_B \varepsilon^{\mu\nu\lambda\rho} A^V_\nu
  F^V_{\lambda\rho} \,,
\label{Bardeen}
\end{align}
with the vector and axial currents,
\begin{align}
  &j'{}^{\mu} \equiv j^\mu + j_B^\mu\,,\qquad j'{}^{\mu}_{\!\!5\,\,}
  \equiv j^\mu_5 + j_{5,B}^\mu\,,
\end{align}
we obtain the anomaly equations
\begin{align}
  &\partial_\mu j'{}^{\mu} = -\left( \frac{C^{122}}{2} + c_B \right) F^V_{\alpha\beta}\tilde F^{A\,\alpha\beta}\,,\\
  &\partial_\mu j'{}^{\mu}_{\!\!5\,\,} =
  -\frac{C^{111}}{4}F^A_{\alpha\beta}\tilde F^{A\,\alpha\beta} -\left(
    \frac{C^{122}}{4} - c_B \right) F^V_{\alpha\beta}\tilde
  F^{V\,\alpha\beta}\,.\nonumber
\end{align}
The electric current $j^{\prime\mu}$ is conserved if $c_B=-C^{122}/2$.
Setting $C^{111}=C^{122} \equiv C/3$, the hydrodynamic Eqs.~(\ref{hydroeqn}) become
\begin{align}
  \partial_\mu T^{\mu\nu} &=  F^{V \nu\lambda} j'_{\lambda}
  +  F^{A \nu\lambda} j'_{5\lambda} \,, \nonumber \\
  \partial_\mu j'{}^{\mu} &= 0 \,,\nonumber\\
  \partial_\mu j'{}^{\mu}_{\!\!5\,\,} &= C  E \cdot  B + (C/3 )E_5 \cdot  B_5  \,.
\end{align}

Using the derivative expansion
\begin{align}
  j'{}^{\mu} & = \rho u^\mu + \kappa_\omega\omega^\mu + \kappa_B
  B^{\mu} + \kappa_{5,B} B_5^{\mu} \,,
\end{align}
where $\kappa_\omega \equiv \xi^2_\omega$, $\kappa_B \equiv
\xi_B^{22}$ and $\kappa_{5,B} \equiv \xi_B^{21}$, we obtain from
(\ref{xi}) and (\ref{xiB}) the conductivities ($\mu_5\equiv \mu^1$,
$\mu\equiv\mu^2$)
\begin{align}
  \kappa_\omega &= 2 C \mu_5 \left( \mu - \displaystyle\frac{\rho}{\epsilon + P_T} \left[\mu^2+\frac{\mu^2_5}{3} \right] \right),\nonumber\\
  \quad \kappa_B &= C \mu_5 \left( 1 - \displaystyle\frac{\mu \rho}{\epsilon + P_T} \right) ,\nonumber\\
  \kappa_{5,B} &= C \mu \left( 1 - \displaystyle\frac{1}{2}\frac{\mu
      \rho}{\epsilon + P_T}\left[1+\frac{\mu_5^2}{3\mu^2}\right]
  \right) .
\end{align}
There are analogous transport coefficients in the axial current
$j^\mu_5$ \cite{Kirsch11}.  The axial fields $E_{5\mu}$ and $B_{5\mu}$
are not needed and can now be switched off.
The first term in $\kappa_B$ and $\kappa_\omega$, $\kappa_B =
C\mu_5$ and $\kappa_\omega=2C\mu\mu_5$, is the leading term in the {\em chiral
  magnetic} (CME) \cite{Kharzeev, Fukushima} and {\em chiral vortical
  effect} \cite{CVE}, respectively.\footnote{$\kappa_{5,B}$ represents
another effect, which we added for completeness, but it seems not to be realized
in heavy-ion collisions.}  They are in agreement with those found in
the isotropic case \cite{Isachenkov, Pu, Kirsch11}. The second term
proportional to  $\rho/(\epsilon+P_T)$
actually depends on the dynamics of the fluid\footnote{In \cite{Zakharov}
this term was considered as a one-loop correction in an effective theory and $(\epsilon+P)/\rho$
was interpreted as the corresponding  infrared cutoff in the energy/momentum
integration.}  and therefore on $\varepsilon_p$.

The dependence of $\kappa_B$ on $\varepsilon_p$ can be made more
visible by introducing an average pressure $\bar P=(2P_T+P_L)/3$ such that
$\epsilon = 3 \bar P$.
Assuming $\varepsilon_p$ to be small (see Fig.~\ref{epsplot}), we expand the CME-coefficient $\kappa_B$
 to linear order in $\varepsilon_p$,
\begin{align}\label{result}
 \kappa_B & \approx C \mu_5 \left( 1 - \displaystyle\frac{\mu \rho}{\epsilon + \bar P} \left[ 1-\frac{\varepsilon_p}{6}\right] \right) \,.
\end{align}
At freeze-out the elliptic flow coefficient $v_2 \approx \varepsilon_p/2$ \cite{Kolb}.
For small momentum anisotropies, the CME thus increases linearly in $v_2$.

\section{Fluid-gravity model}\label{sec3}

In this section we construct the gravity dual of a static anisotropic
plasma with diagonal stress-energy momentum $T_{\mu\nu}={\rm
  diag}(\epsilon, P_T, P_T, P_L)$ and charge densities $\rho^a$.

We start from a five-dimensional $U(1)^n$ Einstein-Maxwell theory
in an asymptotic AdS space. The action is
\begin{align}
  S&=\frac{1}{16\pi G_5} \int d^5x \sqrt{-g} \left[ R - 2\Lambda
    - F^a_{MN} F^{aMN}  \right.  \\
  &\qquad\qquad\qquad~~ \left. + \frac{S_{abc}}{6 \sqrt{-g}}
    \varepsilon^{PKLMN} A^a_P F^b_{KL} F^c_{MN} \right] \,, \nonumber
\end{align}
where $\Lambda=-6$ is the cosmological constant. As usual, the $U(1)$
field strengths are defined by
\begin{align}
  F^a_{MN} &= \partial_M A^a_N - \partial_N A^a_M \,,
\end{align}
where $M,N,... = 0,...,4$ and $a=1,...,n$. The Chern-Simons term
$A\wedge F \wedge F$ encodes the information of the triangle anomalies
in the field theory \cite{Son}. In fact, the Chern-Simons coefficients
$S_{abc}$ are related to the anomaly coefficients $C_{abc}$ by
\begin{align}
  C_{abc} = S_{abc}/(4\pi G_5) \label{CSrel} \,.
\end{align}

The corresponding equations of motion are given by the combined system
of Einstein-Maxwell and Maxwell equations,
\begin{align}
  G_{MN} - 6 g_{MN} &= T_{MN}
  \,,\label{Einstein}\\
  \nabla_M F^{aMP} &= -\frac{S_{abc}}{8\sqrt{-g}} \varepsilon^{PMNKL}
  F^b_{MN} F^c_{KL} \,,
\label{Maxwell}
\end{align}
where the energy-momentum tensor $T_{MN}$ is
\begin{align}
  T_{MN} &= - 2 \left( F^a_{MR} F^{aR}{}_N + \frac{1}{4} g_{MN}
    F^a_{SR} F^{aSR} \right) \,.
\end{align}

\subsection{AdS black hole with multiple U(1) charges}

A gravity dual to an {\em isotropic} fluid ($\epsilon=3P$) with
multiple chemical potentials $\mu_a$ ($a=1,...,n$) at finite
temperature $T$ is given by an AdS black hole solution with mass $m$
and multiple $U(1)$ charges $q^a$.  In Eddington-Finkelstein
coordinates, the metric and $U(1)$ gauge fields of this solution are
\begin{align}
  ds^2 &= - f(r) dt^2 + 2 dr dt + r^2 d\vec x^2\,, \nonumber\\
  A^a  &= - A^a_0(r) dt \,, \label{AdSBH}
\end{align}
where
\begin{align}
  f(r)&= r^2 - \frac{m}{r^2} +\sum_a \frac{(q^a)^2}{r^4}\,,\nn\\
  A^a_0(r) &= \mu^a_\infty + \frac{\sqrt{3} q^a}{2 r^2} \,.
\label{param}
\end{align}
The constants $ \mu^a_\infty$ can be fixed such that the gauge fields
vanish at the horizon.  In case of a single charge ($n=1$), the
background reduces to an ordinary Reissner-Nordstr\o m black hole
solution in $AdS_5$ \cite{Chamblin}.

The temperature $T$ and chemical potentials $\mu^a$ of the fluid are
defined by
\begin{align}
  T &=\frac{\kappa}{2\pi} = \frac{f'(r_+)}{4\pi} =
      \frac{2r_+^6 -  \sum_a (q_a)^2}{2\pi r_+^5} \,,\\
  \mu^a &= A^a_0(r_+) - A^a_0(r_\infty) \,,
\end{align}
where $r_+$ is the outer horizon defined by the maximal solution of
$f(r)=0$, and $r_\infty$ indicates the location of the boundary.  The
temperature of the fluid is the Hawking temperature of the black hole
and is computed from the surface gravity $\kappa = \sqrt{\partial_M
  |\chi| \partial^M |\chi|}\vert_{r_+}$, where $|\chi|= (-\chi^M
\chi_M)^{(1/2)}$ is the norm of the timelike Killing vector
$\chi^M=\delta^M_0$ [here $|\chi|=\sqrt{f(r)}$].

\subsection{Anisotropic AdS geometry with multiple U(1) charges}

We now construct a solution for an {\em anisotropic} fluid
($\epsilon=2P_T+P_L$).  An ansatz for an anisotropic AdS black hole
solution is given by
\begin{align}
  ds^2 &= - f(r) dt^2 + 2 dr dt \nn\\&~~~+ r^2 (w_T(r)
  dx^2 + w_T(r) dy^2 + w_{L}(r)dz^2) \,,\nn\\
  A^a &= - A^a_0(r) dt \,.\label{ansatz0th}
\end{align}
The anisotropies are realized via $w_T(r)$ and $w_L(r)$, which are
functions of the momentum anisotropy $\varepsilon_p$ as defined in
(\ref{epsP}),
\begin{align}
  \varepsilon_p=\frac{\langle P_T-P_L \rangle}{\langle P_T+P_L
    \rangle} \,. \label{beta}
\end{align}
In the isotropic case ($\varepsilon_p=0$), these functions are
required to be one, $w_T(r)=w_L(r)=1$, and the background reduces to
the AdS black hole geometry (\ref{AdSBH}).

An analytical solution of the type (\ref{ansatz0th}) is difficult to
find, and we resort to numerics in the next subsection. For this, we need to know the
solution close to the boundary.  An asymptotic solution ($r
\rightarrow \infty$) is given by the four functions
\begin{align}
  A^a_0(r)&= \mu^a_\infty + \frac{\sqrt{3} q^a}{2 r^2} +{\cal O}(r^{-8})  \,,  \nn\\
  f(r)/r^2&=1 - \frac{m}{r^4} +\sum_a \frac{(q^a)^2}{r^6}+{\cal O}(r^{-8})\,, \nn\\
  w_T(r) &= 1 + \frac{w^{(4)}_T}{ r^4} + {\cal O}(r^{-8}) \,,\nn\\
  w_{L}(r) &= 1 + \frac{w^{(4)}_L}{ r^4} + {\cal O}(r^{-8}) \,,
  \label{asympsol}
\end{align}
where $w^{(4)}_L=-2 w^{(4)}_T =-m\zeta/2$, $\mu^a_\infty=const.$, and
$\zeta$ is related to the momentum anisotropy $\varepsilon_p$ by
\begin{align}
\zeta = \frac{2\varepsilon_p}{\varepsilon_p+3} \,. \label{zetabeta}
\end{align}

The functions $w_T(r)$ and $w_L(r)$ have been introduced in view of
the structure of the anisotropic fluid stress-energy tensor. More
precisely, in (\ref{asympsol}) we fixed the $r^{-4}$ coefficients
$w^{(4)}_T$ and $w^{(4)}_L$ such that the fluid stress-energy tensor
is of the diagonal form (\ref{diagTmn}),
$T^{\mu\nu}=\textmd{diag}(\epsilon, P_T, P_T, P_L)$ with
$\epsilon = 2P_T+P_L$. Computing the stress-energy tensor in the
standard way from the asymptotic solution (\ref{asympsol}) via the
extrinsic curvature, see {\em e.g.}\ \cite{boost}, we find the
transverse and longitudinal pressures
\begin{align}
  P_T&=\frac{m-4 w^{(4)}_T-4w^{(4)}_L}{16\pi G_5} =
  \frac{m(1+\zeta)}{16\pi G_5} \,,\\
  P_L&=\frac{m-8 w^{(4)}_T}{16\pi G_5} = \frac{m(1-2\zeta)}{16\pi
    G_5}\,.
\end{align}
Note that if (\ref{zetabeta}) holds true, the pressures $P_T$ and
$P_L$ satisfy (\ref{beta}). Likewise, the charge densities are
\begin{align}
  \rho^a = \frac{\sqrt{3}q^a}{16\pi G_5} \,.
\end{align}
From these relations, we find the useful identity
\begin{align}\label{rhoepsP}
\frac{\rho^a}{\epsilon+P_T} = \frac{\sqrt{3} q^a}{4m(1+\frac{1}{4}\zeta)}\,,
\end{align}
which we will need later.

\subsection*{Numerical solution}

We now use shooting techniques to solve the system of ordinary
differential equations (ODE) which follows from the equations of motion
(\ref{Einstein}) and (\ref{Maxwell}) upon substituting the ansatz
(\ref{ansatz0th}). The idea is to vary the metric and gauge fields at
some minimal value $r_+$ in the radial direction, integrate outwards
and find solutions with the correct asymptotic behavior
(\ref{asympsol}).  A similar method was previously applied
in~\cite{Erdmenger2011}.

We first need to study the asymptotic solution near $r_+$ and near the
boundary at $r_\infty\gg r_+$ (we choose $r_\infty = 50$ in our
numerics). We define $r_+$ by the maximal solution of
\begin{align}
  f(r_+)=0 \label{r+}
\end{align}
and use scale invariance to set $r_+ = 1$.  We then expand the
functions in the metric and gauge fields near $r_+$ in powers of the
parameter $\varepsilon = \frac{r}{r_+}-1 \ll 1$ and substitute them
into the equations of motion. In this way, we find that the only
independent variables are $\{f'(r_+), w_T(r_+), w_L(r_+), w_L'(r_+)\}$
since the gauge field parameters $A^a_0(r_+)$ can be set to zero using
gauge invariance, $A^a_0(r_+) = 0$.  The other parameters at $r_+$ can
be expressed in terms of these four parameters, {\em e.g.}\
$w'_T(r_+)=w_T(r_+) w'_L(r_+)/w_L(r_+)$.

The near-boundary solution is given by (\ref{ansatz0th}) with
(\ref{asympsol}) and is parameterized by the values $(\zeta, m, q^a,
\mu^a_\infty$). The final set of data is summarized in the following
table:
\begin{equation}
\begin{array}{|c | c|}
\hline
r = r_+ = 1 		& r= r_\infty \gg r_+ \nonumber\\\hline
A^a_0(r_+) = 0 		& \mu^a_\infty 	\nonumber\\
f(r_+) = 0		 &	f(r_\infty)  \nonumber\\
f'(r_+) = \mathrm{fixed}		& A^{a\prime}_0(r_\infty) \nonumber\\
w_L(r_+) = \mathrm{var} 		&	w_L(r_\infty) \nonumber\\
w_T(r_+) = \mathrm{var} 		&	w_T(r_\infty) \nonumber\\
w_L'(r_+) = \mathrm{var} 	&	\nonumber\\\hline
\end{array}
\end{equation}
Parameters not listed are related to those in the table by the
equations of motion.

To integrate the equations we proceed as follows.  We fix $\zeta$ and
vary three parameters at $r_+$, namely $w_T(r_+)$, $w_L(r_+) $ and
$w_L'(r_+) $, by choosing a grid with suitable number of sites (in our
case $20^3-40^3$). The value $f'(r_+)$ can be thought of as the
temperature of the system and will simply be fixed to some value. It
turns out that the form of the functions $w_{L,T}(r)$ does not depend
on this parameter.
For each site in the grid we numerically solve the system of ODEs and
determine the pair $(m, q^a)$ from the known asymptotics of
$A^{a\prime}_0(r = r_\infty)$ and $f(r = r_\infty)$. This ensures that
the analytical and numerical values for these quantities coincide.

\begin{figure}[ht]
\centering
\includegraphics[angle=0, width=8.5cm]{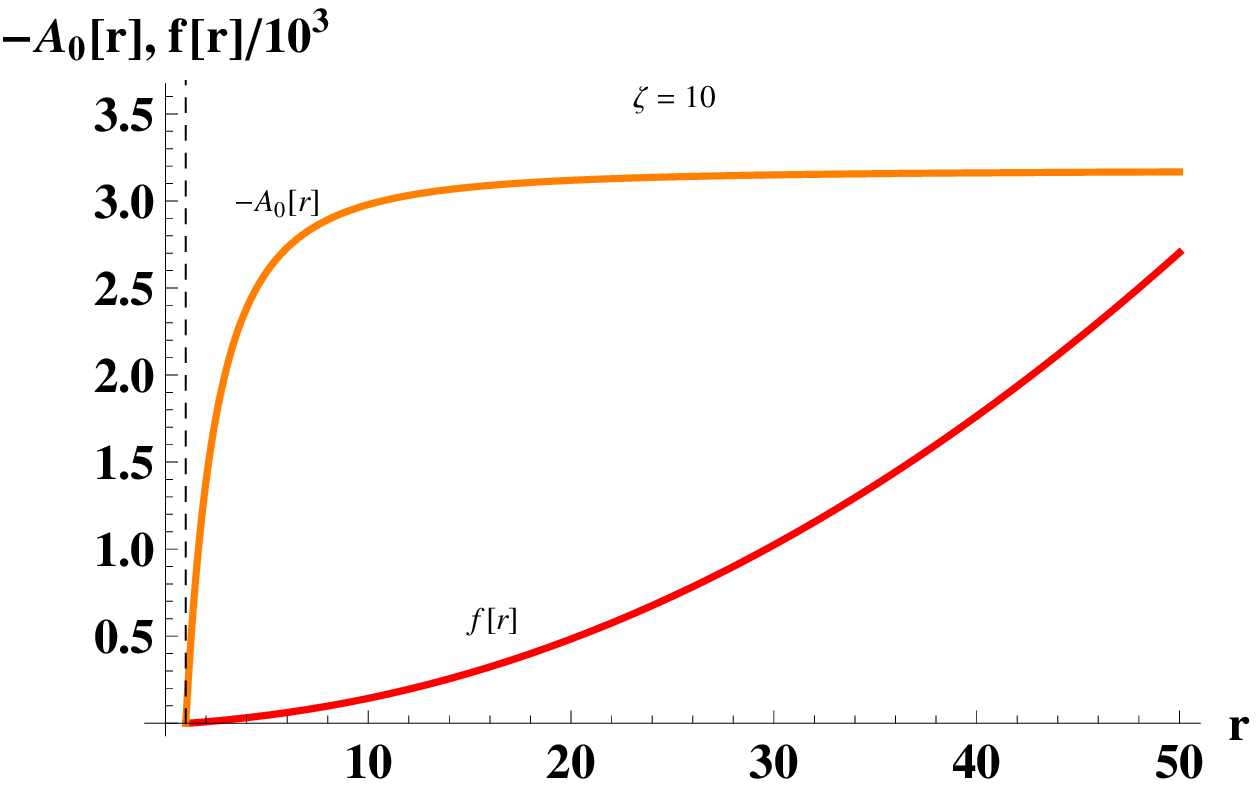}
\includegraphics[angle=0, width=8.5cm]{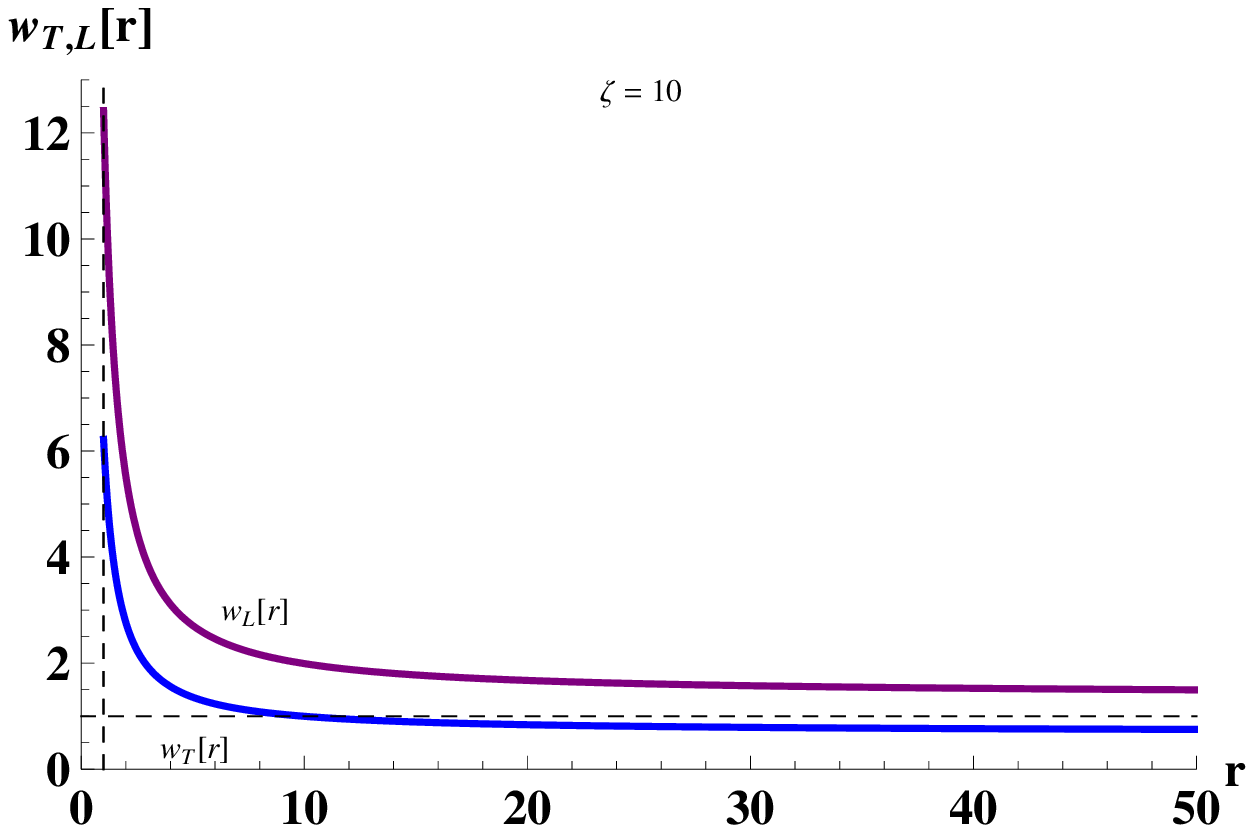}
\caption{\label{fuplot}  Numerical plots of $f(r)$, $A_0(r)$, $w_T(r)$ and $w_L(r)$ for $\zeta=10$
($r_+=1$). We get $w_L(r_+)=12.42$.}
\end{figure}

We then calculate the combined residual
\begin{align}
  & {\rm res}_\infty[w_T(r_+), w_L(r_+), w_L'(r_+))] \nonumber\\
  &= ( w_L^{\#}(r_\infty) - w^*_L(r_\infty))^2 + ( w_T^{\#}(r_\infty) -
  w^*_T(r_\infty))^2,
\end{align}
where $w_{L, T}^{\#}(r_\infty)$ are the numerical values, and $w_{L, T}^{*}(r_\infty)$
are the analytical values given by (\ref{asympsol}).  We
interpolate the residual by a piecewise linear function and find its
global minimum by the simulated annealing method \cite{minimization}.
The result of the minimization is shown in Fig.~\ref{fuplot}, which
depicts numerical plots of $f(r)$, $A_0(r)$, $w_T(r)$ and $w_L(r)$ for
$n=1$.

\begin{figure}[ht]
\centering
\includegraphics[angle=0, width=8.5cm]{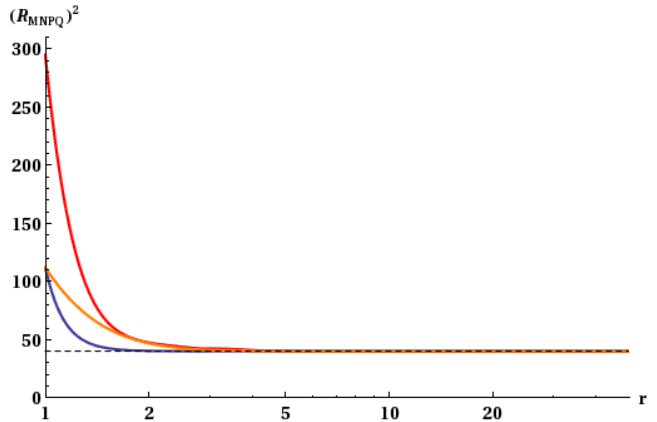}
\caption{\label{R2plot}  Numerical plots of $(R_{MNPQ})^2$ for $\zeta=10$, $q\neq 0$ (red), $\zeta=10$, $q=0$ (orange),
 and $\zeta=0$, $q= 0$ (blue). }
\end{figure}

We conclude this section with a comment on $r_+$. In the isotropic
case, $r_+$ is simply the size of the horizon of the AdS black hole
geometry. For nonvanishing anisotropies and vanishing $U(1)$ charges,
a naked singularity was found at $r_+$ \cite{Witaszczyk}, implying
that the static background does not exist indefinitely. The
singularity is mild in the sense that there is a notion of ingoing
boundary conditions and possible instabilities are absent at the
linear level in the anisotropy parameter \cite{Witaszczyk}.  This
behavior may persist even for nonvanishing $U(1)$ charges, even
though it was difficult to see the singularity in our numerics, {\em cf.}\
Figure~\ref{R2plot}.
Despite this subtlety, we show in the next section that, at least for
small anisotropies where the bulk geometry approximates a black hole
solution, the singular geometry may be used to compute some transport
coefficients of the fluid.

\section{Holographic vortical and magnetic conductivities}\label{sec4}

We will now compute the chiral vortical and magnetic conductivities $\xi^a_\omega$ and
$\xi^{ab}_B$ from first-order corrections to the numerical AdS
geometry (\ref{ansatz0th}) using the fluid-gravity correspondence \cite{Hubeny}.

\subsection{First-order corrected background}

In order to become a dual to a multiply charged fluid, the AdS geometry
(\ref{ansatz0th}) must be boosted along the four-velocity of the fluid
$u_\mu$ ($\mu=0,...,3$). The boosted version of (\ref{ansatz0th}) is
\begin{align}
  ds^2 &= \left(r^2 w_T(r) P_{\mu\nu}-f(r) u_\mu u_\nu\right) dx^\mu dx^\nu -2u_\mu dx^\mu dr \nn\\
  &~~~ - r^2 (w_T(r)- w_L(r)) v_\mu v_\nu dx^\mu dx^\nu \,, \nonumber\\
  A^a &= (A_0^a(r) u_\mu + {\cal A}^a_\mu ) dx^\mu \,,
\label{0thordersol}
\end{align}
where $P^{\mu\nu} = g^{\mu\nu} + u^\mu u^\nu$, and $f(r)$, $A_0^a(r)$, $w_T(r)$ and $w_L(r)$
are numerically known functions.  As in hydrodynamics, the four-vector $v^\mu$ determines
the direction of the longitudinal axis, cf.\ Sec.~2. Following
\cite{Son, Kirsch11}, we have formally introduced constant background
gauge fields ${\cal A}^a_\mu$ to model external electromagnetic
fields, such as the magnetic fields $B^{a\mu}$ needed for the chiral
magnetic effect.

The transport coefficients $\xi^a_\omega$ and $\xi^{ab}_B$ can now be
computed using standard fluid-gravity techniques \cite{Hubeny}. We
closely follow \cite{Son, Yee, Kirsch11}, in which these transport
coefficients were determined for an isotropic fluid with one and three
charges ($n=1,3$).  We work in the static frame $u_\mu=(-1,0,0,0)$,
$v_\mu=(0,0,0,1)$, and consider vanishing background fields ${\cal
  A}^a_\mu$ (at $x^\mu=0$). The transport coefficients $\xi^a_\omega$
and $\xi_B^{ab}$ measure the response of the system to rotation and
the perturbation by an external magnetic field.  We therefore slowly
vary the velocity $u_{\mu}$ and the background fields ${\cal A}^a_\mu$
up to first order as
\begin{align}
  u_{\mu} = (-1, x^{\nu}\partial_{\nu}u_i)\,,\qquad {\cal A}^a_\mu =
  (0, x^{\nu}\partial_{\nu} {\cal A}^a_i) \,.
\label{variations}
\end{align}
We may also vary $m$ and~$q$ in this way, but it turns out that
varying these parameters has no influence on the transport
coefficients $\xi^a_\omega$ and $\xi_B^{ab}$.

Because of the dependence on $x^\mu$, the background (\ref{0thordersol})
is no longer an exact solution of the equations of motion. Instead
with varying para\-meters the solution (\ref{0thordersol}) receives
higher-order corrections, which are in this case of first order in the
derivatives.

An ansatz for the first-order corrected metric and gauge fields is
given by
\begin{align}
  ds^2 &= \lbr -f(r)+\tilde g_{tt}\rbr dt^2 +2 \lbr 1+\tilde g_{tr} \rbr dt dr
  \nonumber\\
  &~~~+ r^2   (w_T(r)   dx^2 + w_T(r) dy^2 + w_{L}(r)dz^2) \nonumber\\
  &~~~+\tilde g_{ij}dx^i dx^j-2 x^{\nu} \partial_\nu u_i dr dx^i \nonumber\\
  &~~~+ 2\lbr \lbr f(r) - r^2 \rbr x^\nu\partial_\nu u_i +
  \tilde g_{ti}\rbr dt dx^i \,, \nn\\
  A^a&=\lbr -A^a_0(r) +\tilde A^a_t\rbr dt \nonumber\\
  &~~~+ \lbr A^a_0(r) x^\nu \partial_\nu u_i + x^\nu \partial_\nu{\cal
    A}^a_i + \tilde A^a_i \rbr dx^i\,, \label{ansatz}
\end{align}
where the first-order corrections are denoted by
\begin{align}
  \tilde g_{MN} = \tilde g_{MN}(r)\,,\quad \tilde A_M^a = \tilde
  A_M^a(r) \,.
\end{align}
As in \cite{Yee}, we work in the gauge
\begin{align}
  \tilde g_{rr}=0\,,\quad \tilde g_{r\mu}\sim u_\mu\,,\quad \tilde
  A^a_r=0\,, \quad \sum_{i=1}^{3}\tilde g_{ii}=0 \,.
\end{align}
The first-order corrections can be obtained by substituting the ansatz
(\ref{ansatz}) into the equations of motion (\ref{Einstein}) and
(\ref{Maxwell}). The computation is straight-forward but lengthy and
has been shifted to Appendix~\ref{appC} [we set
$\mu^a_\infty=A_0^a(r_\infty)=0$ there, see Sec.~\ref{sec2C} for a
discussion]. As a result, we find the following corrections:
\begin{align}
  \tilde g_{tr} &= \tilde g_{tt} = \tilde A^a_t = 0 \,, \nonumber\\
  \tilde g_{ti}(r)&= {f(r)} \int^r_\infty dr'\,
  \frac{ 1}{ w_L(r')^{1/2} r'\left(f(r')\right)^2}\\
  &\times \left(\int_{r_+}^{r'} dr''\, I(r'')-{w_L(r_+)^{1/2} r_+ f'(r_+)}C_i\right)\,,\nonumber\\
  \tilde A^a_i(r) &= \int_\infty^r dr' { \frac{1}{ r' f(r')
      w_L(r')^{1/2}} } \left[ Q_i^a(r') - Q_i^a(r_+) \right.
  \nonumber\\
  &~~~ - \left. C_i {r_+} A_0^a{}'(r_+) w_L(r_+)^{1/2}+ {r' \tilde
      g_{ti}(r')} A_0^a{}'(r') \right] \,,\nonumber
\end{align}
with
\begin{align}
  I(r) &= \sum_{a=1}^n {4} A_0^a{}'(r) \Big(
  Q^{a}_i(r)-Q^{a}_i(r_+)\nonumber\\
  &~~~~~~~~~~~~~~~~~~~~-C_i {r_+} w_L(r_+)^{1/2} A_0^a{}'(r_+) \Big)\,,\nn\\
  Q_a^i &\equiv \frac{1}{2}\CS_{abc} A_0^b A_0^c \epsilon^{ijk}\left(
\partial_j u_k\right)+ \CS_{abc} A_0^b \epsilon^{ijk}\left(\partial_j {\cal A}^c_k\right), \nn \\
C^i &= \frac{ 4 c(r_+) }{w_L(r_+)^{1/2}} \nonumber\\
&~~~\times \Bigg(\frac{1}{3} \CS_{abc}{A_0^a(r_+)A_0^b(r_+)A_0^c(r_+)}\epsilon^{ijk}\left(\partial_j u_k\right)\nonumber\\
&~~~~~~+\frac{1}{2} \CS_{abc}{A_0^a(r_+)A_0^b(r_+) }\epsilon^{ijk}\left(\partial_j {\cal A}^c_k\right)\Bigg)\,, \nonumber\\
c(r_+) &= { { \frac{1}{r_+ (f'(r_+)- {4}
\sum_{a}  A_0^a(r_+) A_0^a{}'(r_+)) }}}\,,\nonumber
\end{align}
and $r_+$ as defined around (\ref{r+}) [$\tilde g_{ij}$ can be
obtained by solving (\ref{E_ij}) in Appendix~\ref{appC} but will not be needed
here].

\subsection{Holographic conductivities}

On the boundary of the asymptotic AdS space (\ref{ansatz}), the metric
and gauge fields couple to the fluid stress-energy tensor and $U(1)$
currents, respectively. Holographic renormalization \cite{Skenderis}
provides relations between these currents and the near-boundary
behavior of their dual bulk fields.  For the magnetic and vortical
effects, we need the $U(1)$ currents $j^{a\mu}$, which are related to
the bulk gauge fields $A^{a\mu}$ by \cite{Skenderis, Yee3}
\begin{align}
  j^{a \mu} &= \lim_{r\rightarrow\infty}\frac{r^2}{8\pi
    G_5}\eta^{\mu\nu} A^a_\nu(r) \,. \label{hr}
\end{align}
Expanding the solution in $\frac{1}{r}$ and substituting only the
corrections $\tilde A^{a}_{\mu}$, we get the currents
\begin{align}
  \tilde j^{a \mu} &= \lim_{r\rightarrow\infty}\frac{r^2}{8\pi G_5}\eta^{\mu\nu}\tilde A^a_\nu(r)  \nonumber\\
  &= \frac{1}{16\pi G_5} \eta^{\mu\nu} \lbr Q_\nu^{a}(r_+) +{r_+ }
  A_0^a{}'(r_+) C_\nu \rbr\,.\label{currentgravity}
\end{align}

\begin{figure}
\centering
\includegraphics[angle=-90, width=8.5cm]{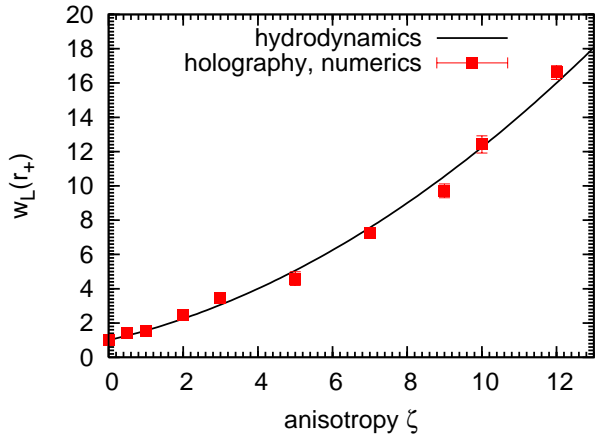}
\caption{\label{zetaplot} Values of $w_L(r_+)$ as a function of the
  anisotropy $\zeta$.  The numerically determined values for
  $w_L(r_+)$ lie on the solid curve, which represents the function
  $(1+\frac{1}{4}\zeta)^2$.}
\end{figure}

Note that, in the isotropic case ($w_L=1$, $P_T=P_L=P$), the prefactor
of the second term of (\ref{currentgravity}) is simply
\begin{align}
  r_+ A_0^{a\prime}(r_+) c(r_+) = \frac{\sqrt{3}}{4m} q^a \,,
  \label{isoprefactor}
\end{align}
as can be seen by substituting the Reissner-Nordstr\o m solution
(\ref{param}) into the left-hand-side of this equation. In the
anisotropic case, we need to show that
\begin{align}
  r_+ A_0^{a\prime}(r_+) c(r_+) \cdot { w_L(r_+)^{-1/2}} =
  \frac{\sqrt{3}q^a}{4m} \cdot \frac{1 }{ 1+\textstyle\frac{1}{4}\zeta
  } \,, \label{anisoprefactor}
\end{align}
which, by (\ref{rhoepsP}), is equivalent to ${\rho^a}/({\epsilon +
  P_T})$. This equation holds in particular if the first and second
factors on both sides agree individually. The first factors correspond
to (\ref{isoprefactor}), which is expected to hold, at least
approximately for small anisotropies $\zeta$. The second factors are
identical if $w_L(r_+, \zeta)= (1+\frac{1}{4} \zeta)^2$. We find
numerically (for $n=1$) that $w_L(r_+)$ indeed satisfies this
equation, see Fig.~\ref{zetaplot}. Thus (\ref{anisoprefactor}) holds
numerically, at least in the limit of small~$\zeta$.

Comparing (\ref{currentgravity}) with the general expansion
\begin{align}
  \tilde j^{a \mu}&= \xi^a_\omega\, \omega^\mu + \xi_B^{a b}\,B^{b\mu}
 \nonumber\\
  &= \xi^a_\omega \,\textstyle
  \half\epsilon^{\nu\rho\sigma\mu}u_\nu\partial_\rho u_\sigma +
  \xi_B^{a b} \epsilon^{\nu\rho\sigma\mu}u_\nu\partial_\rho {\cal
    A}^b_\sigma\, ,  \label{current_hydrodynamics}
\end{align}
we finally obtain the coefficients
\begin{align}
  \xi^a_\omega&= \frac{4}{16 \pi G_5}\Bigg( \CS^{abc} {\mu^b\mu^c}
  - \frac{2}{3} \frac{\rho^a}{\epsilon + P_T} \CS^{bcd} \mu^b\mu^c\mu^d \Bigg)
  \label{xi_hol}\,,\\
  \xi^{ab}_B &= \frac{4}{ 16 \pi G_5}\Bigg( \CS^{abc} {\mu^c} -
  \frac{1}{2}\frac{\rho^a}{\epsilon + P_T} \CS^{bcd} \mu^c\mu^d
  \Bigg)\,,\label{xiB_hol}
\end{align}
with $\mu^a\equiv A^a_0(r_+)$ [since $A^a_0(\infty)=0$].  Using the relation (\ref{CSrel}),
we find that the holographically computed transport coefficients
(\ref{xi_hol}) and (\ref{xiB_hol}) coincide exactly with those found
in hydrodynamics, (\ref{xi}) and (\ref{xiB}).

\subsection{Subtleties in holographic descriptions of the CME}

The conservation of the electromagnetic current requires the
introduction of the Bardeen counter\-term into the action.  In AdS/QCD
models of the CME, this typically leads to a vanishing result for the
electromagnetic current \cite{Rubakov, Rebhan}. The problem is related
to the difficulty of introducing a chemical potential conjugated to a
nonconserved chiral charge \cite{Rubakov, Landsteiner}.  It is
possible to modify the action to obtain a conserved chiral
charge~\cite{Rubakov}.  This charge is however only gauge-invariant
when integrated over all space in homogeneous configurations.

In AdS black hole models of the CME, one usually introduces a chiral
chemical potential dual to a gauge-invariant current, despite it being
anomalous \cite{Landsteiner, Kirsch11}. The prize to pay is
the appearance of a singular bulk gauge field at the horizon, a
phenomenon which seems to be generic in AdS black hole models of the
CME.

Careful holographic renormalization shows that, in the presence of
Chern-Simons terms, there is an additional term on the right-hand side
of (\ref{hr}) \cite{Yee3}. This term is of the form
\begin{align}
  &\hat j_a^\mu = - \frac{\CS_{abc}}{8 \pi G_5}
  \epsilon^{\mu\nu\rho\sigma}A_{b \nu}^{(0)}(x)\partial_\rho A_{c
    \sigma}^{(0)}(x)
\label{jhat}
\,,
\end{align}
where $A_{a \mu}^{(0)}(x)$ are the 0th-order coefficients in a
$\frac{1}{r}$ expansion of the bulk
gauge fields $A_{a \mu}(r,x)$. In (\ref{variations}) we expanded the
background gauge fields ${\cal A}^a_\mu$ around zero and set
${A}_\nu^{a(0)}= \mu^a_\infty u_\nu=0$. This allowed us to ignore
terms in (\ref{hr}) coming from (\ref{jhat}) (at least to first order in the
derivatives).

Problems arise if $\mu_\infty^a \neq 0$. To see this, let us restrict
again to two charges ($n=2$) as in Sec.~\ref{sec2C} and define
axial and vector gauge fields by $A_{\mu}^{A}=A_{\mu}^{1}$ and
$A_{\mu}^{V}=A_{\mu}^{2}$.  Then $\hat j^\mu = \hat j_2^\mu$ gives
rise to additional contributions of the type
\begin{align}
\hat j^{\mu}  \supset \varepsilon^{\mu\nu\rho\sigma}
{A}_\nu^{A(0)}(x) {F}^{V(0)}_{\rho\sigma}(x) \,,
\end{align}
which are forbidden by electromagnetic gauge invariance
\cite{Rubakov}, unless ${A}_\nu^{A(0)}(x)=0$. However, in general
${A}_\nu^{A(0)}(x)= \mu_5^\infty u_\nu$ (at \mbox{$x=0$}) with some
constant $\mu_5^\infty$.  We should thus set $\mu_5^\infty=0$ [Note
that this does not imply $\mu_5=A^A_0(r_\infty)-A^A_0(r_+)=0$].  This
corresponds to a nonvanishing gauge field at the horizon, as noticed
also in \cite{Landsteiner, Kirsch11}.

\section{Conclusions}

Our main result is (\ref{result}), which gives the chiral magnetic conductivity $\kappa_B$
for an anisotropic plasma. It explicitly shows the dependence on the momentum anisotropy~$\varepsilon_p$.
We also computed the CME coefficient in the holographic
dual model and found numerical agreement with the hydrodynamic result for small anisotropies.

\begin{acknowledgments}
We thank Johanna Erdmenger, Patrick Kerner and Andreas Sch\"afer for discussions
and helpful comments. I.~K. is grateful to Berndt M\"uller for pointing out Ref.~\cite{Wang}
and Quan Wang for answering questions
related to the measurement of the charge separation as a function of $v_2$.
I.~G.\ would like to thank Shi Pu for email correspondence.
\end{acknowledgments}

\appendix
\section*{APPENDIX}

\section{Computation of \texorpdfstring{$\partial_\mu \omega^\mu$}{TEXT}  and
\texorpdfstring{$\partial_\mu B^\mu$}{TEXT} } \label{appA}

In the following we will use the identities
\begin{eqnarray}
  u^{\mu}u^{\lambda}\partial_{\mu}\omega_{\lambda} & = & -\frac{1}{2}\partial_{\mu}\omega^{\mu}\,, \label{ident1} \\
  u^{\mu}u^{\lambda}\partial_{\mu}B_{\lambda} & = & \partial_{\mu}B^{\mu}+2\omega^{\rho}E_{\rho}\, \label{ident2}.
\end{eqnarray}
To find an explicit expression for $\partial_\mu \omega^\mu$, we
compute the term $\omega_\nu \partial_\mu T^{\mu \nu}$ in two ways.
First, using the hydrodynamic equations, we get
\begin{eqnarray}
  \omega_\nu \partial_\mu T^{\mu \nu}=\omega_\nu F^{\nu \mu}j_\mu=\rho \omega_\nu  F^{\nu \mu} u_\mu=\rho\omega_\nu E^\nu \,.
\end{eqnarray}
Next, substituting the stress-energy tensor (\ref{AT}) in this
expression, we find
\begin{align} \label{omegaT}
&\omega_\nu \partial_\mu T^{\mu \nu} \nonumber\\
&~~~ =  (\epsilon+P_T)u^\mu \omega_\nu \partial_\mu u^\nu+\omega_\nu g^{\mu \nu} \partial_\mu P_T-\Delta \omega_\nu v^\mu \partial_\mu v^\nu \nonumber\\
&~~~~~~- v^\nu \omega_\nu v^\mu \partial_\mu \Delta-\Delta v^\nu \omega_\nu \partial_\mu v^\mu   \nonumber \\ \nonumber
&~~~ =  -(\epsilon+P_T) u^\mu u^\nu \partial_\mu \omega_\nu +\omega^\mu \partial_\mu P_T-\Delta \omega_\nu \partial^\nu \ln \tau \nonumber\\
&~~~~~~-v^\nu \omega_\nu v^\mu \partial_\mu \Delta-\Delta v^\nu \omega_\nu \partial_\mu v^\mu \,.
\end{align}
Using the identity (\ref{ident1}), we find
\begin{align}
\partial_{\mu}\omega^{\mu} & =  -\frac{2}{\epsilon+P_T}\omega^\mu(\partial_{\mu}P_T-\Delta \partial_\mu \ln \tau -\rho E_\mu\nonumber\\
&\qquad\qquad~~~~~~~~- v_\mu v^\nu \partial_\nu \Delta - \Delta v_\mu \partial_\nu v^\nu )\,. \label{domega}
\end{align}

Similar manipulations of the term $B_\nu \partial_\mu T^{\mu \nu}$ lead to
\begin{align}
  B_\nu \partial_\mu T^{\mu \nu}&=B_\nu F^{\nu \mu} j_\mu=\rho B_\mu E^\mu \,,\label{BT} \\
  B_\nu \partial_\mu T^{\mu \nu} &= -(\epsilon+P_T)u^\mu u^\nu\partial_\mu B_\nu+B^\mu\partial_\mu P_T  \nonumber\\
  &~~~-\Delta B_\nu v^\mu \partial_\mu v^\nu-B_\nu v^\nu v^\mu \partial_\mu \Delta- \Delta B_\nu v^\nu \partial_\mu v^\mu \nonumber\\
  &= -(\epsilon+P_T)(\partial_\mu B^\mu-2 \omega^\mu E_\mu)-\Delta B_\mu \partial^\mu \ln \tau  \nonumber\\
  &~~~-B_\nu v^\nu v^\mu \partial_\mu \Delta- \Delta B_\nu v^\nu
  \partial_\mu v^\mu \label{BT2}\,,
\end{align}
where we used (\ref{ident2}).  From (\ref{BT}) and (\ref{BT2}) we
obtain the following expression:
\begin{align}
  \partial_{\mu}B^{\mu}& =  -2\omega^{\mu}E_{\mu}-\frac{B^\mu}{\epsilon+P_T}(\partial_{\mu}P_T-\Delta \partial_\mu \ln \tau -\rho E_\mu  \nonumber\\
  &\qquad\qquad\qquad~~~~~~- v_\mu v^\nu \partial_\nu \Delta - \Delta
  v_\mu \partial_\nu v^\nu )\,. \label{dB}
\end{align}

The last two terms in (\ref{domega}) and (\ref{dB}) vanish provided
the fluid satisfies
\begin{align}
 \partial_\mu v^\mu=0 \,,\qquad  v^\mu \partial_\mu \Delta=0 \,.
\end{align}
Then  (\ref{domega}) and (\ref{dB}) become identical to
the expressions in (\ref{Paromega}).

\section{Computation of the transport coefficients \texorpdfstring{$\xi_\omega$}{TEXT} and \texorpdfstring{$\xi_B$}{TEXT} }\label{appB}

In this appendix we compute the conductivities $\xi_\omega $ and
$\xi_B$ by solving the system of Eqs.~(\ref{eq1})-(\ref{eq4}).
Following~\cite{Son}, we change variables from $\ln \tau$, $\mu$, $T$
to $\ln \tau$, $\bar{\mu}=\mu/T$ and $P_T$. From (\ref{thermodyn1})
and (\ref{thermodyn2}), we derive the thermodynamic expressions
\begin{align}
\label{Ther1} \left(\frac{\partial \bar{\mu}}{\partial T}\right)_{P_T,\, \ln \tau} & =  -\frac{\epsilon+P_T}{\rho T^2} \,, \\
\label{Ther2} \left(\frac{\partial P_T}{\partial T}\right)_{\bar{\mu}, \,\ln \tau} & =  \frac{\epsilon+P_T}{T}\,, \\
\label{Ther3} \left( \frac{\partial \ln \tau}{\partial T} \right)_{\bar \mu, \, P_T} &=- \frac{1}{\Delta} \frac{\epsilon + P_T}{T}\,.
\end{align}
Using
\begin{align}
 \partial_\mu D &= \frac{\partial D}{\partial P_T}\partial_\mu P_T+\frac{\partial D}{\partial \bar{\mu}}\partial_\mu \bar{\mu}+\frac{\partial D}{\partial\ln \tau}\partial_\mu\ln \tau\,,\\
 \partial_\mu D_B &= \frac{\partial D_B}{\partial P_T}\partial_\mu P_T+\frac{\partial D_B}{\partial \bar{\mu}}\partial_\mu \bar{\mu}+\frac{\partial D_B}{\partial\ln \tau}\partial_\mu\ln
\tau\,,
\end{align}
the first two equations, (\ref{eq1}) and(\ref{eq2}), can be rewritten as
\begin{align}
\label{parxi}   -\xi_\omega + \frac{\partial D}{\partial \bar{\mu}} & = 0\,, &
                -\xi_B + \frac{\partial D_B}{\partial \bar{\mu}} & =  0\,, &\\
               \frac{\partial D}{\partial P_T}-\frac{2 D}{\epsilon+P_T} & = 0\,, &
\label{parDPT}   \frac{\partial D_B}{\partial P_T}-\frac{D_B}{\epsilon+P_T} & = 0\,, & \\
               \frac{\partial D}{\partial\ln \tau}+\frac{2 \Delta D}{\epsilon+P_T}  & = 0\,, &
\label{parDln}   \dfrac{\partial D_B}{\partial\ln \tau}+\dfrac{\Delta D_B}{\epsilon+P_T} & = 0\,. &
\end{align}
Note that (\ref{parDPT}) and (\ref{parDln}) are related by the
thermodynamic identities (\ref{Ther2}) and (\ref{Ther3}).  Using the
ansatz
\begin{align}
 D = T^2d(\bar{\mu},\ln \tau) \,, \quad
 D_B = T d_B(\bar{\mu},\ln \tau) \,,
\end{align}
and (\ref{Ther1}), we obtain two differential equations from
(\ref{eq3}) and (\ref{eq4}),
\begin{align} \nonumber
0 & =  \frac{2 \rho D}{\epsilon+P_T}-2D_B+\frac{\xi_\omega}{T}\\
  & =  T\left(\partial_{\bar{\mu}} d(\bar{\mu},\ln \tau)-2 d_B(\bar{\mu},\ln \tau))\right) \,, \\  \nonumber
0 & =  \frac{ \rho D_B}{\epsilon+P_T}+\frac{\xi_B}{T}-C\bar{\mu} \\
  & =  \partial_{\bar{\mu}}d_B(\bar{\mu},\ln \tau)-C\bar{\mu} \,.
\end{align}
These equations can be integrated to give
\begin{align}
 d_B(\bar\mu,\ln \tau) & = \frac{1}{2} C \bar\mu^2+\beta(\ln \tau) \,, \\
  d(\bar\mu,\ln \tau) & = \frac{1}{3} C \bar\mu^3+ 2 \bar\mu \beta(\ln \tau)+\gamma(\ln \tau)\,,
\end{align}
where $\beta(\ln \tau)$ and $\gamma(\ln \tau)$ are arbitrary functions
of $\ln \tau$. Substituting this back into (\ref{eq3}), (\ref{eq4}),
we get the conductivities
\begin{align}
  \xi_\omega &= C \left(\mu^2 - \frac{2}{3} \frac{\rho \mu^3}{\epsilon+P_T}\right)+2T^2 \beta(\ln \tau) \nonumber \\
       &~~~-\frac{2 \rho T^3}{\epsilon+P_T}\left(2 \bar{\mu} \beta(\ln \tau)+\gamma(\ln \tau)\right) \,,\nonumber \\
  \xi_B &= C \left(\mu-\frac{1}{2}\frac{\rho \mu^2}{\epsilon+P_T}\right)
- \frac{T^2}{\epsilon+P_T}\beta(\ln \tau) \,.  \label{resxi}
\end{align}
The function $\gamma(\ln \tau)$ is forbidden by CPT invariance \cite{Bhattacharya:2011tra}.

\section{First-order corrected background geometry}\label{appC}

In this appendix we compute the first-order corrections to the
background (\ref{0thordersol}) using the ansatz (\ref{ansatz}). The
computation follows that for the three-charge STU model \cite{Cvetic}
presented in \cite{Yee} and \cite{Kirsch11}.

We begin by substituting the ansatz (\ref{ansatz}) into the equations
of motion (\ref{Einstein}) and (\ref{Maxwell}).  We denote the
resulting Maxwell equations, Eqs.~(\ref{Maxwell}) by $M^a_N$ ($a=1,...,n$)
and the components of the Einstein equation, Eqn.~(\ref{Einstein}) by
$E_{MN}$ $M,N=0,...,4$ [$x^M=(t,x^1,x^2,x^3,r)$].  Then, from
$g^{rt}E_{ti} + g^{rr}E_{ri}=0$, we find $\partial_t u_i =0$, and
$E_{tt}$, $E_{rt}$, $E_{rr}$, $E_{tt}$, $M^a_{t}$, and $M^a_{r}$ are
solved by
\begin{align}
\partial_i u_i = \tilde g_{tr} = \tilde g_{tt} = \tilde A^a_t = 0 \,.
\end{align}
The remaining equations are $E_{ij}$, $E_{ti}$, $M^a_{i}$.

From $E_{ij}$ we get
\begin{align}
 -\partial_r\left(r^3 f(r)\partial_r\left( \frac{\tilde g_{ij}(r)}{ r^2} \right)\right)
= 3 r^2 (\partial_{i} u_{j}+ \partial_{j} u_{i}) \label{E_ij}\,.
\end{align}

From $E_{ti}$ we get
\begin{align}
  &\left[ \frac{f'(r)}{f(r)}\left(\frac{2}{r} + \frac{w_T'(r)}{w_T(r)}\right) +\frac{4}{3 f(r)}\left(\sum\limits_{a=1}^n A_0^a{}'(r)^2 -6\right) \right]\!\tilde g_{ti}(r)\nonumber\\
  &~~+ \left(\frac{1}{r} + \ddd\frac{w_L'(r)}{2 w_L(r)} \right) \tilde
  g_{ti}'(r) + \tilde g_{ti}''(r) = 4\sum\limits_{a=1}^n A_0^a{}'(r)
  \tilde A_i^a{}'(r) \label{E_ti}\,,
\end{align}
where a prime denotes the partial derivative $\partial_r$ with respect
to $r$.

From $M^a_{i}$ we get
\begin{align}
  &\partial_r \left[ {w_L(r)^{1/2}} r \left({ f(r)} \tilde A^a_i{}' -
      { \tilde g_{ti}(r)} A_0^a{}'
    \right)\right]\nonumber\\
  &~~~=\partial_r \left( \frac{1}{2}\CS_{abc} A_0^b A_0^c
    \epsilon^{ijk}\left(
      \partial_j u_k\right)+ \CS_{abc} A_0^b
    \epsilon^{ijk}\left(\partial_j
      {\cal A}^c_k\right)\right)\label{M_i} \nonumber\\
  &~~~ \equiv \partial_r Q_i^a(r)\,.
\end{align}

Equation (\ref{E_ij}) depends only on $\tilde g_{ij}(r)$ and can easily be
solved.  The integration of (\ref{M_i}) leads to
\begin{align}\label{solEij}
  & {w_L(r)^{1/2}}\left( {r f(r)} \tilde A^a_i{}'(r) - {r \tilde
      g_{ti}(r)} A_0^a{}'(r) \right)
  \nn\\
  &~~~ = Q_i^a(r) + C_i^a \,.
\end{align}
Here $C_i^a$ are some integration constants, which can be fixed as
\begin{align}
  C_i^a = - Q_i^a(r_+) - C_i {w_L(r_+)^{1/2}} r_+ A_0^a{}'(r_+) \,,
\end{align}
with $r_+$ as in (\ref{r+}) and $C_i =\tilde g_{ti}(r_+)$. This can be
solved for $\tilde A^a_i(r)$,
\begin{align}
  \tilde A^a_i(r) &= \int_\infty^r dr' \frac{ 1}{ r' f(r')
    w_L(r')^{1/2}} \Big[ Q_i^a(r') - Q_i^a(r_+)
  \nonumber\\
  &~~~ -  C_i {r_+} A_0^a{}'(r_+) w_L(r_+)^{1/2}+ {r' \tilde
      g_{ti}(r')} A_0^a{}'(r') \Big] \,.\label{solA}
\end{align}

We still need to determine the constants $C_i$.  Using (\ref{solEij}),
we replace $\tilde A^a_i{}'$ in (\ref{E_ti}) and obtain

\begin{align}
  \left[ \frac{f'(r)}{f(r)}\left(\frac{2}{r} + \frac{w_T'(r)}{w_T(r)}\right) -\frac{8}{3 f(r)}\left(\sum\limits_{a=1}^n A_0^a{}'(r)^2 +3\right) \right]\tilde g_{ti}(r)\nonumber\\
  + \left(\frac{1}{r} + \ddd\frac{w_L'(r)}{2 w_L(r)} \right) \tilde
  g_{ti}'(r) + \tilde g_{ti}''(r) = \frac{1}{w_L(r)^{1/2} r
    f(r)}I(r)\,,\label{Etinew}
\end{align}
where
\begin{align}
I(r) &= \sum_{a=1}^n {4} A_0^a{}'(r) \Big(
Q^{a}_i(r)-Q^{a}_i(r_+)\nonumber\\
&~~~~~~~~
-C_i {r_+} w_L(r_+)^{1/2} A_0^a{}'(r_+) \Big)\,.
\end{align}
A homogeneous solution of this equation $\tilde g_{ti}(r) =
g^{(0)}_{tt}(r) = f(r)$ can be generated by the infinitesimal
coordinate transformation
\begin{align}
dt &\rightarrow dt - \epsilon (dx + dy + dz)\,, \quad dz \rightarrow dz + \epsilon \ddd\frac{dr}{r^2 w_L} \,,\nonumber\\
dx &\rightarrow dz + \epsilon \ddd\frac{dr}{r^2 w_T}\,, \quad dy \rightarrow dy + \epsilon \ddd\frac{dr}{r^2 w_T}\,.
\end{align}
Then, using this homogeneous solution and Appendix~\ref{appD} [$P(r)=f(r)$ and $E(r)=r w_L(r)^{1/2}$ there],
we bring (\ref{Etinew}) to the integrable form
\begin{align}
\partial_r\left(w_L(r)^{1/2} r f^2(r) \partial_r\left( \frac{\tilde g_{ti}(r)}{ f(r)}\right)\right) = I(r)\,.
\end{align}
Solving this equation for $\tilde g_{ti}(r)$ and fixing the integration  constants at $r_+$, we get
\begin{align}
\tilde g_{ti}(r)&=
{f(r)} \int^r_\infty dr'\,
\frac{ 1 }{ w_L(r')^{1/2} r'\left(f(r')\right)^2 } \Big(\int_{r_+}^{r'} dr''\, I(r'')
 \nonumber\\
&~~~-{w_L(r_+)^{1/2} r_+ f'(r_+)}C_i\Big)\,.\label{g_ti solution}
\end{align}

In the Landau frame we require $u_\mu \tau^{\mu\nu}=0$, which in
particular implies the absence of corrections to $T^{ti}$.
Holographic renormalization \cite{Skenderis} translates this into a
constraint for the $r^{-2}$ coefficient of $\tilde g_{ti}(r)$ which is
proportional to the first correction of $T^{ti}$,
\begin{align}
\lim\limits_{r \to \infty} r^2 \, \tilde g_{ti}(r) = 0\,.
\end{align}
In the limit $r \to \infty$, we have the asymptotics
\begin{align}
  f(r) &= O(r^2)\,,\qquad w_L(r) = O(1)\,, \nonumber\\
  &\int_{r_+}^{r} dr'\, I(r') = O(1)\,,
\end{align}
and, from the vanishing of the $r^{-2}$-coefficient of $\tilde
g_{ti}(r)$, we obtain the following equation for $C_i$:
\begin{align} \label{equation}
\ddd{w_L(r_+)^{1/2} r_+ f'(r_+)}C_i = \int_{r_+}^{\infty} dr'\,I(r')\equiv \mathcal{I}_1 + \mathcal{I}_2\cdot C_i\,,
\end{align}
where we defined the integrals
\begin{align}
\mathcal{I}_1 &\equiv {4}\ddd\int_{r_+}^{\infty} dr'\,\sum_{a=1}^n
A_0^a{}'(r')\left(Q^a_i(r')-Q^a_i(r_+)\right) \nonumber\\
 &= \frac{4}{ 3}S_{abc}{A_0^a(r_+)A_0^b(r_+)A_0^c(r_+)} \epsilon^{ijk}\left(\partial_j u_k\right)\nonumber\\
 &~~~~~~~~~ + 2\, \CS_{abc}{A_0^a(r_+)A_0^b(r_+) }\epsilon^{ijk}\left(\partial_j {\cal A}^c_k\right)
\end{align}
and
\begin{align}
\mathcal{I}_2 &\equiv {4} \ddd \int_{r_+}^{\infty} dr'\,\sum_{a=1}^n
A_0^a{}'(r')
\left(- w_L(r_+)^{1/2} r_+ A_0^a{}'(r_+) \right) \nn\\
&= {4}\, w_L(r_+)^{1/2} r_+ \sum_{a=1}^n A_0^a(r_+) A_0^a{}'(r_+) \,.
\end{align}
Solving this for $C_i$, we eventually get
\begin{align}
C^i &= { { \frac{4}{r_+ (f'(r_+)- {4}
\sum_{a}  A_0^a(r_+) A_0^a{}'(r_+)) }}} \cdot \frac{1}{w_L(r_+)^{1/2}} \nonumber\\
&~~~\times \Bigg( \frac{1}{ 3} \CS_{abc}{A_0^a(r_+)A_0^b(r_+)A_0^c(r_+)}\epsilon^{ijk}\left(\partial_j u_k\right)\nonumber\\
&~~~~~~+\frac{1}{2} \CS_{abc}{A_0^a(r_+)A_0^b(r_+) }\epsilon^{ijk}\left(\partial_j {\cal A}^c_k\right)\Bigg)\,.
\end{align}

\section{Integrable form of a linear ordinary differential equation}\label{appD}

In this appendix we present a method to bring an arbitrary linear
ODE of second order to an integrable
form. Let us consider a general form of this equation
\begin{align}
G(g'', g', g, r) \equiv g''(r) + a(r)g'(r) + b(r)g(r) = c(r).\label{ODE}
\end{align}
If we know a homogeneous solution $P(r)$ of this equation, {\em i.e.}\
\begin{align}
 G(P'', P' ,P ,r) = 0\,,
\end{align}
then we can make the substitution
\begin{align}
 g(r) \rightarrow P(r) Q(r), \quad Q'(r) \rightarrow u(r)\label{substitution}
\end{align}
and lower the order of the differential operator (\ref{ODE})
\begin{align}
  G &= P(r)\left(u'(r) + \left[a(r) + 2 \ddd\frac{P'(r)}{P(r)}
    \right]u(r)\right)
  \nonumber\\
  &\equiv P(r) (u'(r) + F(r)u(r))\,.
\end{align}
The term in the brackets can be represented as
\begin{align}
  u'(r) + F(r)u(r) = \frac{1}{A(r)}\partial_r \left(A(r)u(r)\right)\,,
\end{align}
where
\begin{align}
  A(r) = \exp\left\{\int F(r) dr\right\} = P(r)^2\, \exp\left\{\int
    a(r) dr\right\}\,.
\end{align}
Taking into account (\ref{substitution}), we finally bring (\ref{ODE})
to the following integrable form
\begin{align}
  \ddd\frac{1}{P(r)E(r)}\partial_r\left( P(r)^2
    E(r)\partial_r\left(\frac{g(r)}{P(r)}\right) \right) = c(r)\,.
\end{align}
where we defined
\begin{align}
  E(r) \equiv \exp\left\{\int a(r) dr\right\}\,.
\end{align}


\begin{thebibliography}{99}


\bibitem{STAR}
  B.~I.~Abelev {\it et al.}  [STAR Collaboration],
  Phys.\ Rev.\ Lett.\  {\bf 103}, 251601 (2009)
  [\arXiv{arXiv:0909.1739}] $\bullet$
  Phys.\ Rev.\  C {\bf 81}, 054908 (2010)
  [\arXiv{arXiv:0909.1717}] $\bullet$
  D.~Gangadharan [STAR Collaboration],
  J.\ Phys.\ G G {\bf 38} (2011) 124166

\bibitem{PHENIX}
  N.~N.~ Ajitanand, S. Esumi, R.~A.~ Lacey [PHENIX Collaboration], in: Proc. of the RBRC
Workshops, vol. 96, 2010,
\url{http://quark.phy.bnl.gov/~kharzeev/cpodd/}

\bibitem{ALICE}
P.~Christakoglou,
  J.\ Phys.\ G G {\bf 38} (2011) 124165
  [\arXiv{arXiv:1106.2826}]



\bibitem{Kharzeev}
  D.~Kharzeev,
  Phys.\ Lett.\  B {\bf 633}, 260 (2006)
  [\href{http://arxiv.org/abs/hep-ph/0406125}{arXiv:0406125}] $\bullet$
  D.~Kharzeev and A.~Zhitnitsky,
  Nucl.\ Phys.\  A {\bf 797}, 67 (2007)
  [\arXiv{arXiv:0706.1026}] $\bullet$
  D.~E.~Kharzeev, L.~D.~McLerran and H.~J.~Warringa,
  Nucl.\ Phys.\  A {\bf 803}, 227 (2008)
  [\arXiv{arXiv:0711.0950}].

\bibitem{Fukushima}
  K.~Fukushima, D.~E.~Kharzeev and H.~J.~Warringa,
  Phys.\ Rev.\  D {\bf 78}, 074033 (2008)
  [\arXiv{arXiv:0808.3382}].

\bibitem{Vilenkin}
  A.~Vilenkin,
  Phys.\ Rev.\ D {\bf 20} (1979) 1807 $\bullet$
   Phys.\ Rev.\ D {\bf 22} (1980) 3080 $\bullet$
  Phys.\ Rev.\ D {\bf 22} (1980) 3067.

\bibitem{Giovannini:1997gp}
  M.~Giovannini and M.~E.~Shaposhnikov,
  Phys.\ Rev.\ Lett.\  {\bf 80}, 22 (1998)
  [\arXiv{arXiv:hep-ph/9708303}] $\bullet$
  Phys.\ Rev.\  D {\bf 57}, 2186 (1998)
  [\arXiv{arXiv:hep-ph/9710234}].

  \bibitem{condmat}
  A.~Y.~Alekseev, V.~V.~Cheianov and J.~Fr\"ohlich,
  Phys.\ Rev.\ Lett.\  {\bf 81} (1998) 3503
  [\arXiv{arXiv:cond-mat/9803346}] $\bullet$
    A.~Vilenkin,
  Phys.\ Rev.\  B {\bf 25}, 4301 (1982)






\bibitem{lattice1}
  P.~V.~Buividovich, M.~N.~Chernodub, E.~V.~Luschevskaya and M.~I.~Polikarpov,
  Phys.\ Rev.\ D {\bf 80} (2009) 054503
  [\arXiv{arXiv:0907.0494}] $\bullet$
  P.~V.~Buividovich, M.~N.~Chernodub, D.~E.~Kharzeev, T.~Kalaydzhyan, E.~V.~Luschevskaya and M.~I.~Polikarpov,
  Phys.\ Rev.\ Lett.\  {\bf 105} (2010) 132001
  [\arXiv{arXiv:1003.2180}] $\bullet$
  V.~V.~Braguta, P.~V.~Buividovich, T.~Kalaydzhyan, S.~V.~Kuznetsov and M.~I.~Polikarpov,
  PoS LATTICE {\bf 2010} (2010) 190
  [\arXiv{arXiv:1011.3795}].

\bibitem{lattice2}
  M.~Abramczyk, T.~Blum, G.~Petropoulos and R.~Zhou,
  PoS {\bf LAT2009} (2009) 181
  [\arXiv{arXiv:0911.1348}].


\bibitem{lattice3}
  A.~Yamamoto,
  Phys.\ Rev.\ D {\bf 84} (2011) 114504
  [\arXiv{arXiv:1111.4681}] $\bullet$
  Phys.\ Rev.\ Lett.\  {\bf 107} (2011) 031601
  [\arXiv{arXiv:1105.0385}].


\bibitem{Mueller}
  B.~M\"uller and A.~Sch\"afer,
  Phys.\ Rev.\  C {\bf 82}, 057902 (2010)
  [\arXiv{arXiv:1009.1053}].

\bibitem{Wang}
  Q.~Wang,
  arXiv:1205.4638 [nucl-ex].


\bibitem{Son}
  D.~T.~Son and P.~Surowka,
  Phys.\ Rev.\ Lett.\  {\bf 103}, 191601 (2009)
  [\arXiv{arXiv:0906.5044}].

\bibitem{Zakharov}
  A.~V.~Sadofyev, V.~I.~Shevchenko and V.~I.~Zakharov,
  Phys.\ Rev.\ D {\bf 83} (2011) 105025
  [\arXiv{arXiv:1012.1958}].

\bibitem{Isachenkov}
   A.~V.~Sadofyev and M.~V.~Isachenkov,
  Phys.\ Lett.\  B {\bf 697}, 404 (2011)
  [\arXiv{arXiv:1010.1550}].

\bibitem{Pu}
  S.~Pu, J.~h.~Gao and Q.~Wang,
  Phys.\ Rev.\  D {\bf 83}, 094017 (2011)
  [\arXiv{arXiv:1008.2418}] $\bullet$
  J.~-H.~Gao, Z.~-T.~Liang, S.~Pu, Q.~Wang and X.~-N.~Wang,
  [\arXiv{arXiv:1203.0725}].

\bibitem{Oz}
  Y.~Neiman and Y.~Oz,
  JHEP {\bf 1103}, 023 (2011)
  [\arXiv{arXiv:1011.5107}].

\bibitem{Zahed}
  M.~Lublinsky and I.~Zahed,
  Phys.\ Lett.\  B {\bf 684} (2010) 119
  [\arXiv{arXiv:0910.1373}].

\bibitem{Kirsch11}
  T.~Kalaydzhyan and I.~Kirsch,
  Phys.\ Rev.\ Lett.\  {\bf 106} (2011) 211601
  [\arXiv{arXiv:1102.4334}].

\bibitem{Erdmenger}
  J.~Erdmenger, M.~Haack, M.~Kaminski and A.~Yarom,
  JHEP {\bf 0901}, 055 (2009)
  [\arXiv{arXiv:0809.2488}].


\bibitem{Huovinen}
  P.~Huovinen and P.~Petreczky,
  Nucl.\ Phys.\ A {\bf 837}, 26 (2010)
  [\arXiv{arXiv:0912.2541}].


 \bibitem{Ryblewski}
  R.~Ryblewski and W.~Florkowski,
  Phys.\ Rev.\ C {\bf 77}, 064906 (2008)
  [\arXiv{arXiv:0804.2427}].

\bibitem{Florkowski}
  W.~Florkowski,
  Phys.\ Lett.\ B\ {\bf 668}, 32  (2008)
  \href{http://arxiv.org/abs/arXiv:0806.2268}{[arXiv:0806.2268}.

\bibitem{Ryblewski:2011aq}
  R.~Ryblewski and W.~Florkowski,
  Eur.\ Phys.\ J.\ C {\bf 71} (2011) 1761
  [\arXiv{arXiv:1103.1260}].


\bibitem{Kolb}
  P.~F.~Kolb, J.~Sollfrank and U.~W.~Heinz,
  Phys.\ Rev.\ C {\bf 62}, 054909 (2000)
  [hep-ph/0006129].



\bibitem{Cvetic}
  K.~Behrndt, M.~Cvetic and W.~A.~Sabra,
  Nucl.\ Phys.\  B {\bf 553}, 317 (1999)
  [\arXiv{arXiv:hep-th/9810227}].




\bibitem{Yee_CME}
  H.~U.~Yee,
  JHEP {\bf 0911}, 085 (2009)
  [\arXiv{arXiv:0908.4189}].

\bibitem{Zayakin}
  A.~Gorsky, P.~N.~Kopnin and A.~V.~Zayakin,
  Phys.\ Rev.\  D {\bf 83}, 014023 (2011)
  [\arXiv{arXiv:1003.2293}].

\bibitem{Rubakov}
  V.~A.~Rubakov,
  [\arXiv{arXiv:1005.1888}].

\bibitem{Landsteiner}
   A.~Gynther, K.~Landsteiner, F.~Pena-Benitez and A.~Rebhan,
  JHEP {\bf 1102}, 110 (2011)
  [\arXiv{arXiv:1005.2587}].

\bibitem{Rebhan}
  A.~Rebhan, A.~Schmitt and S.~A.~Stricker,
  JHEP {\bf 1001}, 026 (2010)
  [\arXiv{arXiv:0909.4782}].

\bibitem{Brits}
  L.~Brits and J.~Charbonneau,
  Phys.\ Rev.\ D {\bf 83}, 126013 (2011)
  [\arXiv{arXiv:1009.4230}].

\bibitem{Landsteiner2}
  I.~Amado, K.~Landsteiner and F.~Pena-Benitez,
  JHEP {\bf 1105}, 081 (2011)
  [\arXiv{arXiv:1102.4577}].

\bibitem{Lifschytz}
  G.~Lifschytz and M.~Lippert,
  Phys.\ Rev.\  D {\bf 80}, 066005 (2009)
  [\arXiv{arXiv:0904.4772}].

\bibitem{Hoyos:2011us}
  C.~Hoyos, T.~Nishioka, A.~O'Bannon,
  [\arXiv{arXiv:1106.4030}].

\bibitem{Bhattacharya:2011tra}
  J.~Bhattacharya, S.~Bhattacharyya, S.~Minwalla and A.~Yarom,
  [\arXiv{arXiv:1105.3733}].

\bibitem{Hu:2011ze}
  Y.~-P.~Hu, P.~Sun and J.~-H.~Zhang,
  Phys.\ Rev.\ D {\bf 83} (2011) 126003
  [arXiv:1103.3773 [hep-th]] $\bullet$
  Y.~-P.~Hu,
  arXiv:1112.4227 [hep-th].





\bibitem{Witaszczyk}
  R.~A.~Janik and P.~Witaszczyk,
  JHEP {\bf 0809}, 026 (2008)
  [\arXiv{arXiv:0806.2141}].

\bibitem{Mateos}
  M.~Chernicoff, D.~Fernandez, D.~Mateos and D.~Trancanelli,
  \arXiv{arXiv:1202.3696} $\bullet$
  D.~Mateos and D.~Trancanelli,
  Phys.\ Rev.\ Lett.\  {\bf 107}, 101601 (2011)
  [\arXiv{arXiv:1105.3472}] $\bullet$
  D.~Mateos and D.~Trancanelli,
  JHEP {\bf 1107}, 054 (2011)
  [\arXiv{arXiv:1106.1637}].

\bibitem{Erdmenger2011}
  J.~Erdmenger, P.~Kerner and H.~Zeller,
  JHEP {\bf 1201}, 059 (2012)
  [\arXiv{arXiv:1110.0007}].







\bibitem{Hubeny}
  S.~Bhattacharyya, V.~E.~Hubeny, S.~Minwalla and M.~Rangamani,
  JHEP {\bf 0802}, 045 (2008)
  [\arXiv{arXiv:0712.2456}].

\bibitem{Rebhan:2011ke}
  A.~Rebhan and D.~Steineder,
  JHEP {\bf 1108} (2011) 153
  [arXiv:1106.3539 [hep-th]].

\bibitem{Rebhan:2011vd}
  A.~Rebhan and D.~Steineder,
  Phys.\ Rev.\ Lett.\  {\bf 108} (2012) 021601
  [arXiv:1110.6825 [hep-th]].

\bibitem{Giataganas:2012zy}
  D.~Giataganas,
  arXiv:1202.4436 [hep-th].



\bibitem{Landsteiner:2011cp}
  K.~Landsteiner, E.~Megias and F.~Pena-Benitez,
  Phys.\ Rev.\ Lett.\  {\bf 107}, 021601 (2011)
  [arXiv:1103.5006 [hep-ph]].

\bibitem{CVE}
  D.~E.~Kharzeev and D.~T.~Son,
  Phys.\ Rev.\ Lett.\  {\bf 106}, 062301 (2011)
  [\arXiv{arXiv:1010.0038}].

\bibitem{Chamblin}
  A.~Chamblin, R.~Emparan, C.~V.~Johnson and R.~C.~Myers,
  Phys.\ Rev.\ D {\bf 60}, 104026 (1999)
  [\arXiv{arXiv:hep-th/9904197}].

\bibitem{boost}
  T.~Kalaydzhyan and I.~Kirsch,
  JHEP {\bf 1102}, 053 (2011)
  [\arXiv{arXiv:1012.1966}].

 \bibitem{minimization}
  S.~Kirkpatrick, C.~D.~Gelatt and M.~P.~Vecchi,
  Science {\bf 220} (1983) 671 $\bullet$
  V.~Cerny,
  Print-82-0540 (COMENIUS).


\bibitem{Yee}
  M.~Torabian and H.~U.~Yee,
  JHEP {\bf 0908}, 020 (2009)
  [\arXiv{arXiv:0903.4894}].



\bibitem{Skenderis}
  M.~Bianchi, D.~Z.~Freedman and K.~Skenderis,
  Nucl.\ Phys.\  B {\bf 631}, 159 (2002)
  [\arXiv{arXiv:hep-th/0112119}].

\bibitem{Yee3}
  B.~Sahoo and H.~U.~Yee,
  JHEP {\bf 1011}, 095 (2010)
  [\arXiv{arXiv:1004.3541}].

\end{thebibliography}
\end{document}